\documentclass[journal]{IEEEtran}

\usepackage{amsmath,amssymb,bm,mathtools}
\usepackage{amsthm}
\usepackage{graphicx}
\usepackage{grffile}

\usepackage{caption}
\usepackage{subcaption}
\usepackage{cite}

\theoremstyle{definition}
\newtheorem{theorem}{Theorem}[section]

\newtheorem{corollary}{Corollary}[section]
\newtheorem{remark}{Remark}[section]

\newtheorem{assumption}{Assumption}[section]

\usepackage{algpseudocode,algorithm}

\newcommand{\argmin}{\operatornamewithlimits{\mathrm{arg\ min}}}
\newcommand{\argmax}{\operatornamewithlimits{\mathrm{arg\ max}}}
\newcommand{\plim}{\operatornamewithlimits{\mathrm{plim}}}
\newcommand{\calS}{\mathcal{S}}
\newcommand{\sign}{\mathrm{sign}}
\newcommand{\txb}{\text{b}}
\newcommand{\txc}{\text{c}}

\newcommand{\txr}{\text{r}}
\newcommand{\txu}{\text{u}}
\newcommand{\txv}{\text{v}}
\newcommand{\txw}{\text{w}}

\newcommand{\txG}{\text{G}}
\newcommand{\txR}{\text{R}}
\newcommand{\txX}{\text{X}}
\newcommand{\Res}{\mathrm{Res}}
\newcommand{\prox}{\mathrm{prox}}
\newcommand{\env}{\mathrm{env}}
\newcommand{\prop}{\mathrm{prop}}
\newcommand{\Pto}{\xrightarrow{\mathrm{P}}}
\newcommand{\abs}[1]{\left\lvert#1\right\rvert}
\newcommand{\norm}[1]{\left\lVert#1\right\rVert}
\newcommand{\Ex}[1]{\mathrm{E}\left[#1\right]}
\newcommand{\paren}[1]{\left(#1\right)}
\newcommand{\sqbra}[1]{\left[#1\right]}
\newcommand{\curbra}[1]{\left\{#1\right\}}

\allowdisplaybreaks[1]

\begin{document}
%--------
% Title
%--------
\title{Noise Variance Estimation Using Asymptotic Residual in Compressed Sensing}
%
%-----------
% Author
%-----------
\author{
	Ryo~Hayakawa% <-this % stops a space
	\thanks{
		This work was supported in part by JSPS KAKENHI Grant Number JP20K23324.
	}%
	\thanks{
		Ryo Hayakawa is with Institute of Engineering, Tokyo University of Agriculture and Technology, Tokyo, Japan (e-mail: hayakawa@go.tuat.ac.jp). 
	}% <-this % stops a space
}%
% The paper headers
\markboth{ACCEPTED TO APSIPA TRANSACTIONS ON SIGNAL AND INFORMATION PROCESSING}%
{R. Hayakawa: Noise Variance Estimation Using Asymptotic Residual in Compressed Sensing}
%
% make the title area
\maketitle
%-------------
% Abstract
%-------------
\begin{abstract}
    In compressed sensing, measurements are typically contaminated by additive noise, and therefore, information about the noise variance is often needed to design algorithms.
    In this paper, we propose a method for estimating the unknown noise variance in compressed sensing problems. 
    The proposed method, called asymptotic residual matching (ARM), estimates the noise variance from a single measurement vector on the basis of the asymptotic result for the $\ell_{1}$ optimization problem. 
    Specifically, we derive the asymptotic residual corresponding to the $\ell_{1}$ optimization and show that it depends on the noise variance. 
    The proposed ARM approach obtains the estimate by comparing the asymptotic residual with the actual one, which can be obtained by empirical reconstruction without the information on the noise variance. 
    For the proposed ARM, we also propose a method to choose a reasonable parameter based on the asymptotic residual. 
    Simulation results show that the proposed noise variance estimation outperforms several conventional methods, especially when the problem size is small. 
    We also show that, by using the proposed method, we can tune the regularization parameter of the $\ell_{1}$ optimization to achieve good reconstruction performance, even when the noise variance is unknown.
\end{abstract}
\IEEEpeerreviewmaketitle
%
%---------------
% Introduction
%---------------
\section{Introduction} \label{sec:intro}
Compressed sensing~\cite{candes2005,candes2006,donoho2006,candes2008} has attracted much attention in the field of signal processing~\cite{lustig2007,lustig2008,hayashi2013,choi2017}. 
One of the main purposes of compressed sensing is to solve underdetermined linear inverse problems of an unknown vector with a structure such as sparsity. 
Although the underdetermined problem has an infinite number of solutions in general, we can often reconstruct the unknown vector by using the sparsity as the prior knowledge appropriately. 
A similar idea can be applied to the reconstruction of other non-sparse structured vectors, e.g., discrete-valued vectors~\cite{aissa-el-bey2015,nagahara2015}, which often appear in wireless communication systems~\cite{choi2017a,sasahara2017,hayakawa2017a}. 

There are various algorithms proposed for compressed sensing. 
In greedy algorithms such as matching pursuit (MP)~\cite{mallat1993} and orthogonal matching pursuit (OMP)~\cite{pati1993,tropp2007}, we update the support of the estimate of the unknown sparse vector in an iterative manner. 
Another approach based on message passing, e.g., approximated belief propagation (BP)~\cite{kabashima2003} and approximate message passing (AMP)~\cite{donoho2009,bayati2011}, utilizes a Bayesian framework for the reconstruction of the structured vector. 
Such message passing-based methods can achieve good reconstruction performance with low complexity for large-scale problems. 
For small-scale problems with a few hundred unknown variables, however, their performance degrades and the algorithms may even diverge. 

Various convex optimization-based approaches have also been studied in the literature on compressed sensing. 
The most popular convex optimization problem for compressed sensing is $\ell_{1}$ optimization (a.k.a. least absolute shrinkage and selection operator (LASSO)~\cite{tibshirani1996}), where the $\ell_{1}$ norm is used as the regularizer to promote the sparsity. 
The iterative shrinkage thresholding algorithm (ISTA)~\cite{daubechies2004,combettes2005,figueiredo2007} and the fast iterative shrinkage thresholding algorithm (FISTA)~\cite{beck2009} can solve the $\ell_{1}$ optimization problem with feasible computational complexity. 
Another promising algorithm is the alternating direction method of multipliers (ADMM)~\cite{gabay1976,eckstein1992,combettes2011,boyd2011}, which can be applied to a wider class of optimization problems than ISTA and FISTA. 
Such optimization-based approaches can also be applied to the reconstruction of other non-sparse structured vectors~\cite{bach2012,hayakawa2018b,fosson2019,hayakawa2019}. 
Unlike the message passing-based methods, the convex optimization-based algorithms converge to the solution of the corresponding optimization problem even for small-scale reconstruction problems. 

The measurement vector in compressed sensing is usually contaminated by additive noise in practice. 
In the design of the algorithms for compressed sensing, information on the noise variance is often required to obtain good reconstruction performance. 
In optimization-based approaches, for example, the objective function and/or the constraint in the problem usually include some parameters to be fixed in advance. 
Since the appropriate value of the parameter depends on the noise variance in general, its information is essential to tune the parameter of the optimization problem, with few exceptions such as square-root LASSO~\cite{belloni2011a,shen2016}. 
If the noise variance and the distribution of the unknown vector are known, we can obtain the optimal regularization parameter in terms of the asymptotic mean squared error (MSE) under several assumptions by using some analytical results~\cite{donoho2009,bayati2011,thrampoulidis2018}. 
Hence, when the noise variance is unknown, we need to estimate it from the measurement vector before the reconstruction of the unknown vector. 

Although several estimation methods for the noise variance have been proposed in the context of linear regression in statistics~\cite{reid2016,dicker2016,liu2020}, some of them mainly consider non-structured vectors and do not exploit the sparsity of the unknown vector. 
On the other hand, some sparsity-based methods such as~\cite{sun2012a,yu2019} cannot be extended to the case with other non-sparse structured vectors in a trivial manner. 
A possible exception is AMP-LASSO~\cite{bayati2013}, which is based on the asymptotic analysis of the MSE of LASSO~\cite{mousavi2013,mousavi2018}. 
The estimate of the noise variance by AMP-LASSO is consistent and can be simply calculated using the reconstructed vector by LASSO with a fixed regularization parameter. 
For small-scale problems, however, we need to choose an appropriate value of the regularization parameter to obtain a good estimate of the noise variance. 
For more details of related work, see Section~\ref{sec:related}. 

In this paper, we propose a novel estimation method for the noise variance on the basis of the asymptotic analysis for the $\ell_{1}$ optimization. 
The proposed approach, referred to as asymptotic residual matching (ARM), uses the fact that the residual of the estimate obtained by the $\ell_{1}$ optimization can be well predicted under some assumptions when the problem size is sufficiently large. 
By using the convex Gaussian min-max theorem (CGMT)~\cite{thrampoulidis2015d,thrampoulidis2018} and a similar procedure to~\cite{miolane2021}, we derive the asymptotic residual in the large system limit, where the problem size goes to infinity. 
The asymptotic residual depends on the noise variance, whereas the empirical residual can be computed without using the noise variance because we just need to solve the $\ell_{1}$ optimization problem. 
We can thus estimate the noise variance by choosing the value whose corresponding asymptotic residual is the closest to the empirical residual. 
Hence, the proposed noise variance estimation firstly solves the $\ell_{1}$ optimization problem with a fixed regularization parameter and then computes the empirical residual of the reconstructed vector. 
After that, we obtain the noise variance whose corresponding asymptotic residual matches the empirical residual. 

As is the case with other methods such as AMP-LASSO, the estimation performance of the proposed method depends on the value of the regularization parameter. 
We thus propose a parameter initialization method for the proposed ARM on the basis of the asymptotic residual. 
The proposed method enables us to choose a reasonable value of the regularization parameter without the computation of the solution of the $\ell_{1}$ optimization problem. 
To further improve the estimation performance, we also propose the iterative approach, where we iterate the estimation of the noise variance and the update of the regularization parameter. 
Hence, unlike the conventional methods, the proposed method can estimate the noise variance without the manual tuning of the regularization parameter. 
Another advantage of the proposed ARM is that we can easily extend it to the case with other non-sparse structured vectors if the distribution is known. 
In this paper, we consider the reconstruction of binary vector as an example, which appears in some communication systems such as multiple-input multiple-output (MIMO) signal detection~\cite{chockalingam2014, yang2015}. 

Simulation results demonstrate that the proposed method can achieve good estimation performance even when the problem size is small. 
By using the estimate of the noise variance for the choice of the regularization parameter, we can obtain good reconstruction performance in compressed sensing even when the noise variance is unknown. 

The rest of the paper is organized as follows. 
We describe the problem considered in this paper in Section~\ref{sec:problem} and related work in Section~\ref{sec:related}. 
We then provide the analytical results for the residual of the $\ell_{1}$ optimization in Section~\ref{sec:result}. 
In Section~\ref{sec:proposed}, we explain the proposed noise variance estimation method based on the analytical result. 
In Section~\ref{sec:extension}, we discuss the extension of the proposed method and show the example for binary vector reconstruction. 
We demonstrate several simulation results to show the validity of the proposed method in Section~\ref{sec:simulations}. 
Finally, Section~\ref{sec:conclusion} presents some conclusions. 

In this paper, we use the following notations. 
We denote the transpose by $(\cdot)^{\top}$ and the identity matrix by $\bm{I}$. 
For a vector $\bm{a}=\sqbra{a_{1}\, \cdots \, a_{N}}^{\top} \in \mathbb{R}^{N}$, the $\ell_{1}$ norm and the $\ell_{2}$ norm are given by $\norm{\bm{a}}_{1} = \sum_{n=1}^{N} \abs{a_{n}}$ and $\norm{\bm{a}}_{2} = \sqrt{\sum_{n=1}^{N} a_{n}^{2}}$, respectively. 
We denote the number of nonzero elements of $\bm{a}$ by $\norm{\bm{a}}_{0}$. 
$\sign(\cdot)$ denotes the sign function. 
For a lower semicontinuous convex function $\zeta: \mathbb{R}^{N} \to \mathbb{R} \cup \curbra{+\infty}$, we define the proximity operator and the Moreau envelope as 
\begin{align}
    \prox_{\zeta} (\bm{a}) 
    &= 
    \argmin_{\bm{u} \in \mathbb{R}^{N}} \curbra{ \zeta(\bm{u}) + \frac{1}{2} \norm{\bm{u}-\bm{a}}_{2}^{2}}, \\
    \env_{\zeta} (\bm{a}) 
    &= 
    \min_{\bm{u} \in \mathbb{R}^{N}} \curbra{ \zeta(\bm{u}) + \frac{1}{2} \norm{\bm{u}-\bm{a}}_{2}^{2}}, 
\end{align}
respectively. 
The probability density function (PDF) and the cumulative distribution function (CDF) of the standard Gaussian distribution are denoted as $p_{\txG}(\cdot)$ and $P_{\txG}(\cdot)$, respectively. 
When the PDF of the random variable $X$ is given by $p_{\txX}$, we denote $X \sim p_{\txX}$. 
When a sequence of random variables $\curbra{\Theta_{n}}$ ($n=1,2,\dotsc$) converges in probability to $\Theta$, we denote $\Theta_{n} \Pto \Theta$  as $n \to \infty$ or $\plim_{n \to \infty} \Theta_{n} = \Theta$. 
%
%-----------------
%  Problem Settings
%-----------------
\section{Noise Variance Estimation\\in Compressed Sensing} \label{sec:problem}
A standard problem in compressed sensing is the reconstruction of an $N$ dimensional sparse vector $\bm{x} = \sqbra{x_{1}\ \dotsb\ x_{N} }^{\top} \in \mathbb{R}^{N}$ from its linear measurements given by 
\begin{align}
	\bm{y} = \bm{A} \bm{x} + \bm{v} \in \mathbb{R}^{M}, \label{eq:model}
\end{align}
where $\bm{A} \in \mathbb{R}^{M \times N}$ is a known measurement matrix and $\bm{v} \in \mathbb{R}^{M}$ is an additive noise vector. 
We denote the measurement ratio by $\Delta = M / N$. 
In the scenario of compressed sensing, we focus on the underdetermined case with $\Delta < 1$ and utilize the sparsity of $\bm{x}$ as the prior knowledge for the reconstruction. 

One of the most famous convex optimization problems for compressed sensing is the $\ell_{1}$ optimization given by 
\begin{align}
    \hat{\bm{x}} (\lambda) 
    &= 
    \argmin_{\bm{s} \in \mathbb{R}^{N}} 
    \curbra{
        \frac{1}{2} \norm{ \bm{y}-\bm{A}\bm{s} }_{2}^{2} 
        + \lambda f ({\bm{s}}) 
    }, \label{eq:optimization}
\end{align}
where $f(\bm{s}) = \norm{\bm{s}}_{1}$ is the $\ell_{1}$ regularizer to promote the sparsity of the estimate $\hat{\bm{x}}(\lambda)$ of the unknown vector $\bm{x}$. 
The regularization parameter $\lambda$ ($>0$) controls the balance between the data fidelity term $\frac{1}{2} \norm{\bm{y} - \bm{A} \bm{s}}_{2}^{2}$ and the $\ell_{1}$ regularization term $\lambda f({\bm{s}})$. 
Since the $\ell_{1}$ optimization is the convex optimization problem, the sequence converging to the global optimum can be obtained by several convex optimization algorithms~\cite{combettes2005,beck2009,combettes2011,boyd2011}. 

In this paper, we assume that the noise variance $\sigma_{\txv}^{2}$ is unknown, and tackle the problem of estimating $\sigma_{\txv}^{2}$ from the single measurement $\bm{y}$ and the corresponding measurement matrix $\bm{A}$. 
The knowledge of the noise variance $\sigma_{\txv}^{2}$ is important to design the algorithms for compressed sensing. 
For the optimization problems in~\eqref{eq:optimization}, for example, the reconstruction performance largely depends on the parameter $\lambda$ and its appropriate value is different depending on the noise variance. 
In fact, by using the AMP framework or the CGMT framework, the asymptotically optimal parameter minimizing MSE can be obtained under some assumptions when the noise variance is known~\cite{mousavi2018,thrampoulidis2018}. 
Hence, the accurate estimate of the noise variance is significant to achieve good reconstruction performance in convex optimization-based compressed sensing. 
For other approaches, the information on the noise variance would also be helpful to design the algorithm. 
\section{Related Work} \label{sec:related}
In statistics, several estimation methods for the noise variance have been discussed in the context of linear regression~\cite{dicker2014,dicker2016,janson2017}. 
A method using the residual of the ridge regression has been proposed in~\cite{liu2020}, where simulation results show that it outperforms some conventional approaches. 
The signal-to-noise ratio (SNR) estimation method in~\cite{suliman2017} is also based on the analysis of the ridge regularized least squares. 
Although it has good estimation performance when the number of measurements is sufficiently large, the performance degrades for underdetermined problems like compressed sensing. 
Moreover, the above methods mainly focus on the non-structured unknown vectors, and hence they do not take advantage of the sparsity in the estimation. 

Some sparsity-aware methods have also been proposed for the noise variance estimation, e.g., scaled LASSO~\cite{sun2012a} and refitted cross-validation~\cite{fan2012}. 
In~\cite{reid2016}, the authors have compared the performance of several estimators and have concluded that a promising estimator is given by
\begin{align}
    \hat{\sigma}_{\txv}^{2} 
    &= 
    \frac{1}{M - \norm{\hat{\bm{x}}(\lambda)}_{0}} 
    \norm{\bm{y} - \bm{A} \hat{\bm{x}}(\lambda)}_{2}^{2}. \label{eq:SR}
\end{align}
For the estimator in~\eqref{eq:SR}, however, the regularization parameter $\lambda$ significantly affects the estimation performance and the parameter should be carefully selected. 
Although the cross-validation technique can be used for the choice of $\lambda$, it increases the computational cost of the estimation. 
Even if we use several approximation techniques~\cite{obuchi2016,giordano2019,rad2020,stephenson2020}, we need to obtain the estimate $\hat{\bm{x}}(\lambda)$ for various values of $\lambda$ to choose its appropriate value.
Moreover, these sparsity-aware methods only consider the sparse unknown vector, and the extension of other non-sparse structured vectors is not trivial. 

Another LASSO-based method has also been proposed in~\cite{bayati2013} on the basis of the analysis of the AMP algorithm~\cite{donoho2009,bayati2011}. 
The estimate by AMP-LASSO can be written as 
\begin{align}
    \hat{\sigma}_{\txv}^{2} 
    &= 
    \Delta \hat{\tau}^{2} - \hat{R}(\hat{\tau}), \label{eq:AMP-LASSO}
\end{align}
where $\hat{\tau} = \sqrt{N} \norm{\bm{y} - \bm{A} \hat{\bm{x}}(\lambda)}_{2} / (M - \norm{\hat{\bm{x}}(\lambda)}_{0})$ and 
\begin{align}
    \hat{R}(\tau) 
    &= 
    \tau^{2} \paren{\frac{2 \norm{\hat{\bm{x}}(\lambda)}_{0}}{N} - 1} 
    + 
    \frac{N \norm{\bm{A}^{\top} \paren{\bm{y} - \bm{A} \hat{\bm{x}}(\lambda)}}_{2}^{2}}{\paren{M - \norm{\hat{\bm{x}}(\lambda)}_{0}}^{2}}
\end{align}
(under Assumption~\ref{ass:problem} below). 
The estimate in~\eqref{eq:AMP-LASSO} is consistent and hence the noise variance is well predicted in large-scale problems. 
Simulation results in~\cite{bayati2013} show that the estimation performance of AMP-LASSO is better than several conventional methods such as~\cite{sun2012a,fan2012}. 
For small-scale problems, however, we need to choose an appropriate value of the regularization parameter in LASSO to obtain a good estimate of the noise variance. 
Moreover, the extension to the structure other than sparsity is not discussed explicitly. 

Non-asymptotic and asymptotic analyses have been discussed for the residual $\norm{\bm{y} - \bm{A} \hat{\bm{x}}(\lambda)}_{2}^{2}$ of the LASSO problem in~\cite{miolane2021}. 
Moreover, the tuning method for the regularization parameter $\lambda$ has been proposed on the basis of the analysis. 
However, the regularizer other than the $\ell_{1}$ regularizer has not been considered explicitly in the paper. 
Furthermore, for the tuning of the regularization parameter, we need to solve the optimization problem for many values of $\lambda$, which increases the computational cost. 

Although we focus on the case with measurement matrices $\bm{A}$ composed of independent and identically distributed (i.i.d.) Gaussian elements in this paper, the performance analyses for non-i.i.d.\ cases have been discussed in several papers, e.g.,~\cite{celentano2022,bellec2023}. 
Especially, in~\cite{bellec2023}, the out-of-sample error has been analyzed for general regularizers, including the $\ell_{1}$ regularizer and the elastic-net regularizer. 
The noise variance estimation has also been considered in~\cite{bellec2023}, and the estimate is equivalent to that of AMP-LASSO in the case with i.i.d.~Gaussian distribution. 
%
%----------------
%  Asymptotic Results
%----------------
\section{Asymptotic Residual for $\ell_{1}$ Optimization} \label{sec:result}
In this section, by using the CGMT framework~\cite{thrampoulidis2015d,thrampoulidis2018}, we provide an asymptotic result for the $\ell_{1}$ optimization in~\eqref{eq:optimization}, which will be used in the proposed noise variance estimation in Section~\ref{sec:proposed}. 
Although a part of the result can be derived from the general CGMT-based analysis in~\cite{thrampoulidis2018}, we here derive the explicit formula required in the proposed method. 
We characterize the asymptotic property of the residual $\norm{\bm{y} - \bm{A} \hat{\bm{x}}(\lambda)}_{2}^{2}$ in the following. 
It should be noted that a similar analysis has been discussed in~\cite{miolane2021} for the LASSO problem, though we use more general notation for the regularizer in this paper and actually consider a different problem in Section~\ref{sec:extension}. 

In the analysis, we use the following assumption. 
\begin{assumption} \label{ass:problem}
    The unknown vector $\bm{x}$ is composed of i.i.d.\ random variables with a distribution $p_{\txX}(x)$ which have some mean and variance. 
    The measurement matrix $\bm{A}$ is composed of i.i.d.\ Gaussian random variables with zero mean and variance $1/N$. 
    The noise vector $\bm{v}$ is also Gaussian with mean $\bm{0}$ and covariance matrix $\sigma_{\txv}^{2} \bm{I}$. 
\end{assumption}
In Assumption~\ref{ass:problem}, we assume the Gaussian measurement matrix because it is required to apply CGMT in a rigorous manner. 
However, the universality~\cite{bayati2013,panahi2017,oymak2018} of random matrices suggests that the analytical result also holds for other i.i.d.\ measurement matrix. 
In fact, the simulation result in~\cite{atitallah2017} shows that the result of the CGMT-based analysis is valid even for the measurement matrix from Bernoulli or Laplace distribution. 
Hence, it would be possible to utilize our theoretical results for such cases in practice. 

By using the CGMT framework~\cite{thrampoulidis2018}, we provide the asymptotic property of the residual $\norm{\bm{y} - \bm{A} \hat{\bm{x}}(\lambda)}_{2}^{2}$ in the following. 
It should be noted that the standard CGMT-based analysis gives the asymptotic error performance such as MSE, which is different from the residual analyzed here. 
In the theorem, we consider the large system limit $N, M \to \infty$ with the fixed ratio $\Delta = M / N$, which we simply denote as $N \to \infty$ in this paper. 
\begin{theorem} \label{th:main}
    We assume that Assumption~\ref{ass:problem} is satisfied. 
    We also assume that the optimization problem $\min_{\alpha > 0} \max_{\beta \ge 0} F(\alpha, \beta)$ has a unique optimizer $(\alpha^{\ast}, \beta^{\ast})$\footnote{The uniqueness can be proven under some conditions. For example, if the set of minimizers of the problem over $\alpha$ is bounded, the uniqueness of $\alpha^{\ast}$ is guaranteed. However, it would be difficult to eliminate the assumption completely in the general case. For detailed discussions, see~\cite[Remark~19]{thrampoulidis2018}.}, where 
    \begin{align}
        F(\alpha, \beta)
        &= 
        \frac{\alpha \beta \sqrt{\Delta}}{2} 
        + \frac{\sigma_{\txv}^{2} \beta \sqrt{\Delta}}{2 \alpha}
        - \frac{1}{2} \beta^{2} 
        - \frac{\alpha \beta}{2 \sqrt{\Delta}} \notag \\
        &\hspace{4mm}
        + 
        \frac{\beta \sqrt{\Delta}}{\alpha} 
        \Ex{\env_{\frac{\alpha \lambda}{\beta \sqrt{\Delta}} f} \paren{X + \frac{\alpha}{\sqrt{\Delta}} G}} \label{eq:SO}
    \end{align}
    and $X \sim p_{\txX}, G \sim p_{\txG}$. 
    Then, the asymptotic value of the objective function in~\eqref{eq:optimization} and the residual for the optimizer $\hat{\bm{x}}(\lambda)$ are given by
    \begin{align}
        \plim_{N \to \infty}\ 
        \frac{1}{N} 
        \paren{ 
            \frac{1}{2} \norm{ \bm{y} - \bm{A} \hat{\bm{x}}(\lambda) }_{2}^{2} 
            + \lambda f \paren{\hat{\bm{x}}(\lambda)} 
        } 
        &= 
        F\paren{\alpha^{\ast}, \beta^{\ast}}, \label{eq:convergence_objective} \\
        \plim_{N \to \infty}\ 
        \frac{1}{N} \norm{\bm{y} - \bm{A} \hat{\bm{x}}(\lambda)}_{2}^{2} 
        &= 
        (\beta^{\ast})^{2}, \label{eq:convergence_residual}
    \end{align}
    respectively. 
\end{theorem}
\begin{proof}
See Section~\ref{app:proof}. 
\end{proof}

To compute $\alpha^{\ast}$ and $\beta^{\ast}$ in Theorem~\ref{th:main}, we need to optimize the function $F(\alpha, \beta)$ in~\eqref{eq:SO}. 
Fortunately, for some distribution $p_{\txX}(x)$, we can write the expectation $\Ex{\env_{\frac{\alpha \lambda}{\beta \sqrt{\Delta}} f} \paren{X + \frac{\alpha}{\sqrt{\Delta}} G}}$ in~\eqref{eq:SO} with an explicit formula. 
For example, when the distribution of the unknown vector is given by the Bernoulli-Gaussian distribution as 
\begin{align}
    p_{\txX} (x) 
    &= 
    p_{0} \delta_{0}(x) + (1 - p_{0}) p_{\txG}(x), \label{eq:Bernoulli-Gaussian}
\end{align}
the expectation can be easily computed with the PDF and CDF of the standard Gaussian distribution, where $\delta_{0}(\cdot)$ denotes the Dirac delta function and $p_{0} \in (0,1)$. 
For details of the derivation, see Section~\ref{app:exp}. 
For the Bernoulli distribution given by 
\begin{align}
    p_{\txX} (x) 
    &= 
    p_{0} \delta_{0}(x) + (1 - p_{0}) \delta_{0}(x - 1), \label{eq:Bernoulli}
\end{align}
we can also obtain the explicit form of the expectation in a similar way. 
In such case, we can easily optimize $F(\alpha, \beta)$ by line search techniques such as ternary search and golden-section search~\cite{luenberger2008}. 
When the exact computation of the expectation is difficult, we can approximate it by the Monte Carlo method with many realizations of $X$ and $G$. 

\begin{figure}[t]
    \centering
    \includegraphics[width=80mm]{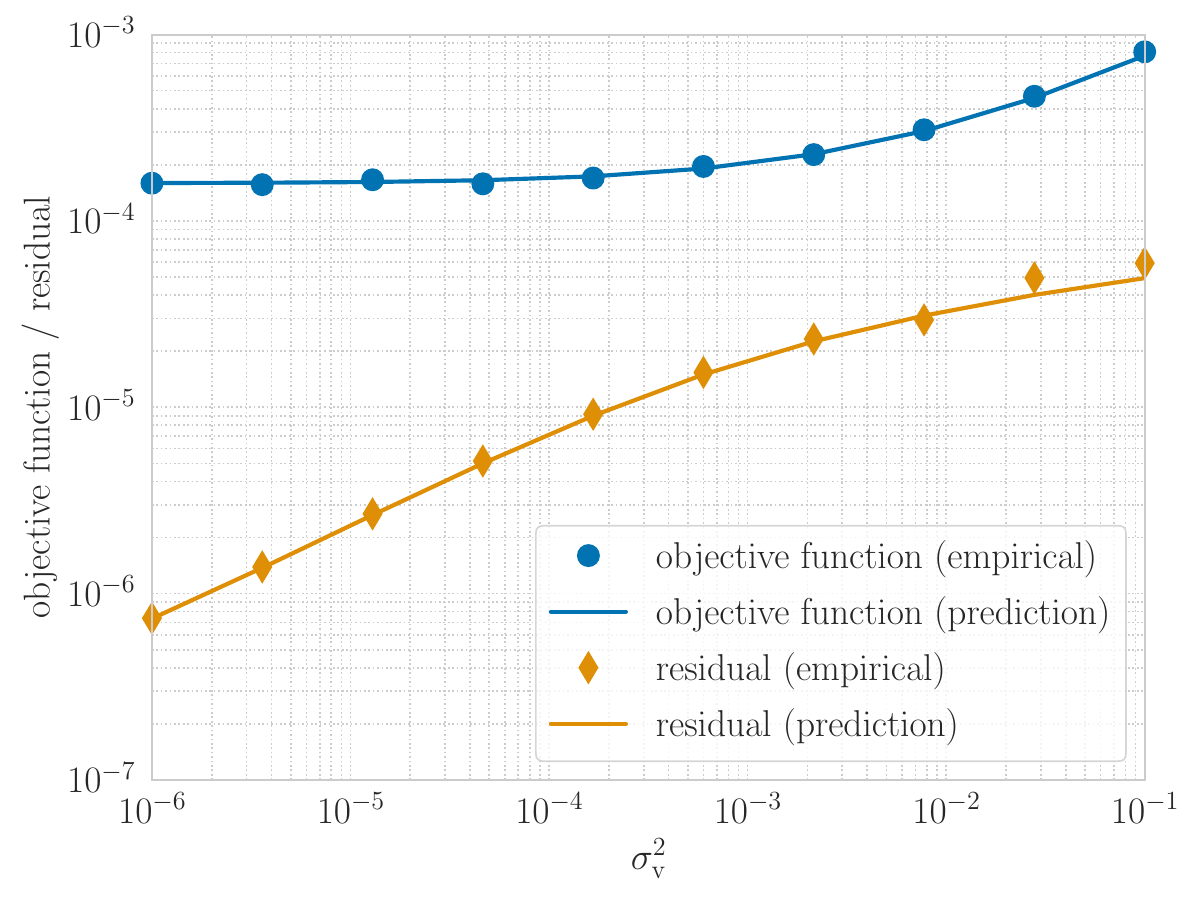}
    \caption{Objective function and residual for the optimizer ($N = 100$, $M = 90$, $p_{0} = 0.8$, $\lambda = 0.001$, $p_{\txX}(x)$: Bernoulli-Gaussian distribution).}
    \label{fig:objective_residual}
\end{figure}

From Theorem~\ref{th:main}, we can predict the optimal value of the objective function and the residual in the empirical reconstruction for compressed sensing problems. 
Figure~\ref{fig:objective_residual} shows the comparison between the empirical values and their prediction, where $N = 100$ and $M = 90$. 
The distribution of the unknown vector $\bm{x}$ is Bernoulli-Gaussian in~\eqref{eq:Bernoulli-Gaussian} with $p_{0} = 0.8$. 
In the figure, `empirical' means the empirical value of the objective function $\frac{1}{N} \paren{\frac{1}{2} \norm{ \bm{y} - \bm{A} \hat{\bm{x}}(\lambda) }_{2}^{2} + \lambda f \paren{\hat{\bm{x}}(\lambda)}}$ and the residual $\frac{1}{N} \norm{\bm{y} - \bm{A} \hat{\bm{x}}(\lambda)}_{2}^{2}$, where $\hat{\bm{x}}(\lambda)$ is obtained by the $\ell_{1}$ optimization in~\eqref{eq:optimization} with $\lambda = 0.001$. 
The empirical results are averaged over $100$ independent trials. 
For the reconstruction, we use the LASSO solver of scikit-learn~\cite{pedregosa2011scikit}. 
In Fig.~\ref{fig:objective_residual}, we also plot the asymptotic value obtained from Theorem~\ref{th:main} as `prediction'. 
We can see that the empirical value agrees well with the theoretical prediction for both the objective function and the residual. 
%
%----------------
%  Proposed Noise Variance Estimation
%----------------
\section{Proposed Noise Variance Estimation} \label{sec:proposed}
In this section, we propose an algorithm for the estimation of the noise variance $\sigma_{\txv}^{2}$ on the basis of the asymptotic analysis in Section~\ref{sec:result}. 
\subsection{Asymptotic Residual Matching}
The proposed method uses the fact that the residual $\Res\paren{\hat{\bm{x}}(\lambda)} \coloneqq \frac{1}{N} \norm{\bm{y} - \bm{A} \hat{\bm{x}}(\lambda)}_{2}^{2}$ can be approximated by $\paren{\beta^{\ast}}^{2}$ from~\eqref{eq:convergence_residual} when $N$ and $M$ are sufficiently large. 
Since the function $F(\alpha, \beta)$ to be optimized depends on the regularization parameter $\lambda$, the noise variance $\sigma_{\txv}^{2}$, and the probability of zero $p_{0}$, the value of the optimal $\beta^{\ast}$ can be considered as a function of $(\lambda, \sigma_{\txv}^{2}, p_{0})$. 
To explicitly show the dependency, we denote $\beta^{\ast}$ as $\beta^{\ast}(\lambda, \sigma_{\txv}^{2}, p_{0})$ hereafter. 
On the other hand, we can calculate the empirical estimate $\hat{\bm{x}}(\lambda)$ and the corresponding residual $\Res\paren{\hat{\bm{x}}(\lambda)}$ from~\eqref{eq:optimization} without using $\sigma_{\txv}^{2}$ in the reconstruction. 
We can thus estimate the noise variance by choosing $\sigma^{2}$ which minimizes the difference $\abs{\beta^{\ast}(\lambda, \sigma^{2}, \hat{p}_{0})^{2} - \Res\paren{\hat{\bm{x}}(\lambda)}}$, where $\hat{p}_{0}$ is the estimate of the probability $p_{0}$. 
Hence, the proposed estimate of the noise variance is given by 
\begin{align}
    \hat{\sigma}_{\txv}^{2} 
    &= 
    \argmin_{\sigma^{2} > 0}\ 
    \abs{
        \beta^{\ast}(\lambda, \sigma^{2}, \hat{p}_{0})^{2} 
        - \Res\paren{\hat{\bm{x}}(\lambda)}
    }. \label{eq:optimization_sigma}
\end{align} 

In the proposed optimization problem~\eqref{eq:optimization_sigma}, we need the estimate of the probability $p_{0}$ when $p_{0}$ is unknown. 
In this paper, we use the rough estimate given by $\hat{p}_{0} = 1 - \norm{\bm{y}}_{2}^{2}/M$ on the basis of~\cite[Eq.~(10)]{lopes2013}, which means that $\frac{N}{M} \norm{\bm{y}}_{2}^{2}$ is an estimate of $\norm{\bm{x}}_{2}^{2}$ for the measurement matrix satisfying Assumption~\ref{ass:problem}. 
For simplicity, we here assume that the second moment of the non-zero value of $X \sim p_{\txX}$ is $1$ as in~\eqref{eq:Bernoulli-Gaussian} and~\eqref{eq:Bernoulli}. 
The problem in~\eqref{eq:optimization_sigma} is a scalar optimization problem over $\sigma^{2}$, and hence the optimal value can be obtained by line search methods such as the ternary search and the golden-section search~\cite{luenberger2008}. 

\begin{remark}[Advantage of Using Residual of $\ell_{1}$ Optimization] \label{rem:ad_residual}
    The proposed estimation method uses the asymptotic result for the residual of the $\ell_{1}$ optimization problem. 
    Although we can use the asymptotic value of the objective function in~\eqref{eq:convergence_objective} for the noise variance estimation, the performance would be worse in that case. 
    This is because the line of the objective function is flat especially when the noise variance $\sigma_{\txv}^{2}$ is small as shown in Fig.~\ref{fig:objective_residual}. 
    The conventional SNR estimation method in~\cite{suliman2017} has the same problem because it utilizes the asymptotic result for the objective function of the ridge regularized least squares. 
    In fact, the simulation results in~\cite{suliman2017} show that the estimation performance becomes worse when the linear system is underdetermined. 
    Another reason for the performance degradation is that the reconstruction performance of the ridge regularized least squares severely degrades for underdetermined problems like compressed sensing. 
    On the other hand, as shown in Fig.~\ref{fig:objective_residual}, the residual of the $\ell_{1}$ optimization decreases more rapidly than the objective function as the noise variance $\sigma_{\txv}^{2}$ decreases. 
    We thus conclude that we should use not the objective function but the residual for the noise variance estimation. 
\end{remark}
\subsection{Initialization of $\lambda$}
\begin{figure}[t]
    \centering
    \includegraphics[width=80mm]{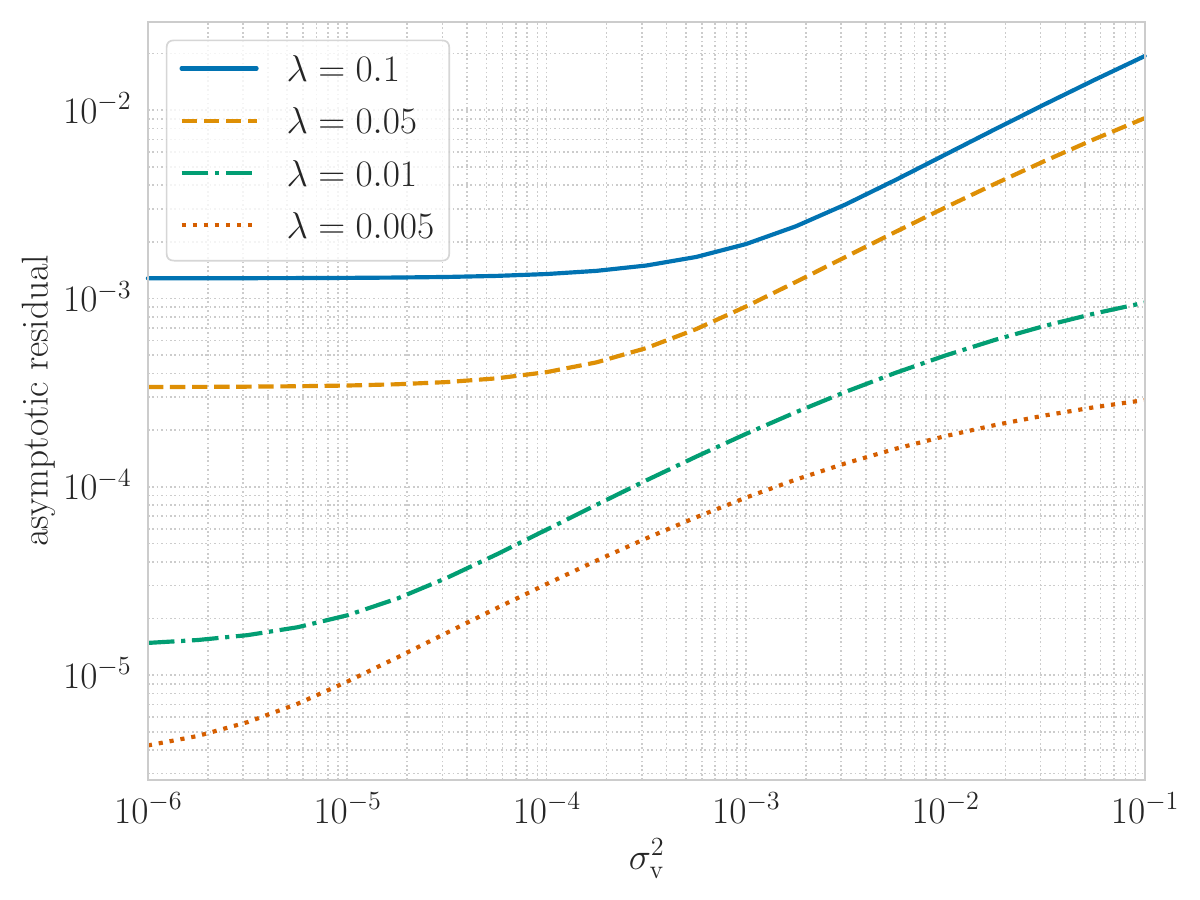}
    \caption{Asymptotic residual of the $\ell_{1}$ optimization ($\Delta = 0.8$, $p_{0} = 0.9$).}
    \label{fig:residual_lambda}
\end{figure}

Since the prediction of the residual from Theorem~\ref{th:main} is not exactly accurate for finite $N$, the estimation performance of the proposed optimization problem~\eqref{eq:optimization_sigma} depends on the parameter $\lambda$. 
Figure~\ref{fig:residual_lambda} shows the asymptotic residual $\beta^{\ast}(\lambda, \sigma^{2}, p_{0})^{2}$ for different values of $\lambda$ when $\Delta = 0.8$ and $p_{0} = 0.9$. 
From the figure, we can see that the slope of the line depends on $\lambda$. 
In the case of Fig.~\ref{fig:residual_lambda}, it is difficult to distinguish the noise variance between $10^{-6}$ and $10^{-4}$ if we use the regularization parameter $\lambda = 0.1$ because the line of the asymptotic residual with $\lambda = 0.1$ is flat in the range. 
Moreover, if the empirical value of the residual is smaller than the line unfortunately, there might be no positive candidate of $\sigma_{\txv}^{2}$. 
From the above discussion, it would be better to use $\lambda = 0.005$ when the true noise variance $\sigma_{\txv}^{2}$ is small. 
On the other hand, when $\sigma_{\txv}^{2} = 10^{-1}$, for example, the choice $\lambda = 0.1$ seems the best of the four in Fig.~\ref{fig:residual_lambda} because it has the steepest slope around $\sigma_{\txv}^{2} = 10^{-1}$. 
We need to choose an appropriate value of $\lambda$ to achieve better estimation performance. 

To tackle this problem, we propose an initialization method based on the max-min approach. 
We define the quantity
\begin{align}
    D(\lambda, \sigma^{2}, \hat{p}_{0}) 
    = 
    \frac{\beta^{\ast}\paren{\lambda, (1 + \varepsilon) \sigma^{2}, \hat{p}_{0}}^{2}}{\beta^{\ast}\paren{\lambda, \sigma^{2}, \hat{p}_{0}}^{2}}, \label{eq:diff}
\end{align}
which represents how much $\beta^{\ast}$ increases when the value of $\sigma^{2}$ increases to $(1 + \varepsilon) \sigma^{2}$ ($\varepsilon > 0$). 
Since the scale of $\beta^{\ast}$ is quite different for different $\lambda$ and $\sigma^{2}$ as shown in Fig.~\ref{fig:residual_lambda}, we take the ratio of $\beta^{\ast}\paren{\lambda, (1 + \varepsilon) \sigma^{2}, \hat{p}_{0}}^{2}$ to $\beta^{\ast}\paren{\lambda, \sigma^{2}, \hat{p}_{0}}^{2}$. 
The larger $D(\lambda, \sigma^{2}, \hat{p}_{0})$ is, the more rapidly $\paren{\beta^{\ast}}^{2}$ increases along with the increase of $\sigma^{2}$. 
Hence, from the discussion of the previous paragraph, $\lambda$ should be chosen so that $D(\lambda, \sigma^{2})$ becomes large. 
Since the noise variance $\sigma^{2}$ is unknown of course, we here adopt the max-min approach to obtain the proper regularization parameter as 
\begin{align}
    \lambda_{\prop} (\hat{p}_{0})
    &= 
    \argmax_{\hspace{2mm} 0 \le \lambda \le \lambda_{\mathrm{max}}}
    \curbra{
        \min_{\sigma^{2} \in \Sigma} D(\lambda, \sigma^{2}, \hat{p}_{0})
    }, \label{eq:lmd_prop}
\end{align}
where $\lambda_{\mathrm{max}}$ restricts the range of $\lambda$ and $\Sigma$ denotes the set of the candidate values for the noise variance, e.g., $\Sigma = \curbra{10^{-5}, 10^{-3}, 10^{-1}}$. 
Note that we do not require that the true noise variance $\sigma_{\txv}^{2}$ is included in $\Sigma$. 
From~\eqref{eq:lmd_prop}, we can choose a reasonable regularization parameter $\lambda$ in the sense that it maximizes $D(\lambda, \sigma^{2}, \hat{p}_{0})$ for the worst $\sigma^{2} \in \Sigma$, without the empirical reconstruction of $\bm{x}$. 
\subsection{Iterative Estimation}
To improve the performance, we also propose an iterative approach as in Algorithm~\ref{alg:proposed}. 
We firstly compute the initial regularization parameter $\lambda_{0} = \lambda_{\prop} (\hat{p}_{0})$ with~\eqref{eq:lmd_prop}, and then iterate the updates of the estimated noise variance $\hat{\sigma}_{\txv, t}^{2}$ and the regularization parameter $\lambda_{t}$, where $t$ denotes the iteration index. 
At the $t$-th iteration, the estimate of the noise variance $\hat{\sigma}_{\txv, t}^{2}$ is calculated by solving~\eqref{eq:optimization_sigma} with $\lambda = \lambda_{t-1}$. 
Using the estimate $\hat{\sigma}_{\txv, t}^{2}$, we update the regularization parameter as 
\begin{align}
    \lambda_{t} 
    &= 
    \argmax_{\hspace{2mm} 0 \le \lambda \le \lambda_{\mathrm{max}}}
    D(\lambda, \hat{\sigma}_{\txv, t}^{2}, \hat{p}_{0}). \label{eq:lambda_update}
\end{align}
to obatin a good regularization parameter for $\hat{\sigma}_{\txv, t}^{2}$. 
If the estimate $\hat{\sigma}_{\txv, t}^{2}$ is closer to the true value $\sigma_{\txv}^{2}$ than $\hat{\sigma}_{\txv, t-1}^{2}$, the new parameter $\lambda_{t}$ is expected to be better than the preivious parameter $\lambda_{t-1}$. 
After $T$ iterations, the proposed ARM in Algorithm~\ref{alg:proposed} outputs the final estimate of the noise variance $\hat{\sigma}_{\txv, T}^{2}$. 
\begin{algorithm}[t]
	\caption{Proposed Asymptotic Residual Matching (ARM)}
	\begin{algorithmic}[1]
		\Require{measurement vector $\bm{y}$, measurement matrix $\bm{A}$, estimated probability $\hat{p}_{0}$}
		\Ensure{estimated noise variance $\hat{\sigma}_{\txv, T}^{2}$}
        \State{${\displaystyle \lambda_{0} = \argmax_{\hspace{2mm} 0 \le \lambda \le \lambda_{\mathrm{max}}} \curbra{ \min_{\sigma^{2} \in \Sigma} D(\lambda, \sigma^{2}, \hat{p}_{0})}}$}
        \For{$t = 1$ to $T$}
            \State{Solve~\eqref{eq:optimization} with $\lambda = \lambda_{t-1}$ and obtain $\hat{\bm{x}}(\lambda_{t-1})$.} 
            \State{${\displaystyle \Res\paren{\hat{\bm{x}}(\lambda_{t-1})} = \frac{1}{N} \norm{\bm{y} - \bm{A} \hat{\bm{x}}(\lambda_{t-1})}_{2}^{2}}$}
            \State{${\displaystyle \hat{\sigma}_{\txv, t}^{2} = \argmin_{\sigma^{2} > 0}\ \abs{\beta^{\ast} \paren{\lambda_{t-1}, \sigma^{2}, \hat{p}_{0}}^{2} - \Res\paren{\hat{\bm{x}}(\lambda_{t-1})} }}$}
            \State{${\displaystyle \lambda_{t} = \argmax_{\hspace{2mm} 0 \le \lambda \le \lambda_{\mathrm{max}}} D \paren{\lambda, \hat{\sigma}_{\txv, t}^{2}, \hat{p}_{0}}}$}
        \EndFor
	\end{algorithmic}
	\label{alg:proposed}
\end{algorithm}
\section{Extension to Other Structured Vectors} \label{sec:extension}
Although we have focused on the reconstruction of the sparse vector in the previous sections, the proposed approach using the asymptotic residual can also be applied to the reconstruction of other non-sparse structured vectors. 
For example, the noise variance estimation with the proposed ARM approach can be utilized in the reconstruction of discrete-valued vectors because the CGMT-based analysis has been applied to the problem~\cite{thrampoulidis2018a, hayakawa2020a}. 
In this section, we mainly describe the noise variance estimation for the binary vector reconstruction as the simplest example. 

In the binary vector reconstruction, we estimate the unknown binary vector $\bm{x}_{\txb} \in \curbra{1, -1}^{N}$ from its linear measurements $\bm{y}_{\txb} = \bm{A} \bm{x}_{\txb} + \bm{v} \in \mathbb{R}^{M}$. 
In this seciton, we consider the unknown vector $\bm{x}_{\txb}$ with the distribution 
\begin{align}
    p_{\txX, \txb} (x) 
    &= 
    \frac{1}{2} \delta_{0}(x + 1) + \frac{1}{2} \delta_{0}(x-1). \label{eq:distribution_binary}
\end{align}
Such problem often appears in several communication systems, such as the MIMO signal detection~\cite{chockalingam2014, yang2015} and the multiuser detection~\cite{verdu1998}. 
As in the sparse vector reconstruction discussed in the previous sections, we require the information on the noise variance to obtain better performance with various methods~\cite{datta2010,datta2012,sasahara2017,hayakawa2017a} for the binary vector reconstruction. 

A simple approach for the binary vector reconstruction is the box relaxation method~\cite{tan2001,yener2002,thrampoulidis2018a}, which solves the optimization problem 
\begin{align}
    \hat{\bm{x}}_{\txb} 
    &= 
    \argmin_{\bm{s} \in \sqbra{-1, 1}^{N}} 
    \curbra{
        \frac{1}{2} \norm{ \bm{y}_{\txb} - \bm{A} \bm{s} }_{2}^{2} 
    }. \label{eq:box_relaxation}
\end{align}
Using the indicator function 
\begin{align}
    f_{\txb} (\bm{s}) 
    &= 
    \begin{cases}
        0 & (\bm{s} \in \sqbra{-1, 1}^{N}) \\
        \infty & (\bm{s} \notin \sqbra{-1, 1}^{N})
    \end{cases},
\end{align}
we can rewrite the box relaxation problem in~\eqref{eq:box_relaxation} as 
\begin{align}
    \hat{\bm{x}}_{\txb} 
    &= 
    \argmin_{\bm{s} \in \mathbb{R}^{N}} 
    \curbra{
        \frac{1}{2} \norm{ \bm{y}_{\txb} - \bm{A}\bm{s} }_{2}^{2} 
        + f_{\txb} (\bm{s})
    }. \label{eq:optimization_binary}
\end{align}

The asymptotic analysis in Theorem~\ref{th:main} can be applied to the optimization problem in~\eqref{eq:optimization_binary}. 
The following corollary shows the result of the analysis, which can be proven in the same way as Theorem~\ref{th:main}. 
\begin{corollary} \label{cor:main_binary}
    We assume that Assumption~\ref{ass:problem} is satisfied. 
    We also assume that the optimization problem $\min_{\alpha > 0} \max_{\beta \ge 0} F_{\txb} (\alpha, \beta)$ has a unique optimizer $(\alpha_{\txb}^{\ast}, \beta_{\txb}^{\ast})$, where 
    \begin{align}
        F_{\txb} (\alpha, \beta)
        &= 
        \frac{\alpha \beta \sqrt{\Delta}}{2} 
        + \frac{\sigma_{\txv}^{2} \beta \sqrt{\Delta}}{2 \alpha}
        - \frac{1}{2} \beta^{2} 
        - \frac{\alpha \beta}{2 \sqrt{\Delta}} \notag \\
        &\hspace{4mm}
        + 
        \frac{\beta \sqrt{\Delta}}{\alpha} 
        \Ex{\env_{\frac{\alpha}{\beta \sqrt{\Delta}} f_{\txb}} \paren{X_{\txb} + \frac{\alpha}{\sqrt{\Delta}} G}} \label{eq:SO_binary}
    \end{align}
    and $X_{\txb} \sim p_{\txX, \txb}, G \sim p_{\txG}$. 
    Then, the asymptotic value of the objective function in~\eqref{eq:box_relaxation} and the residual for the optimizer $\hat{\bm{x}}_{\txb}$ are given by 
    \begin{align}
        &\plim_{N \to \infty}\ 
        \frac{1}{N} 
        \paren{ 
            \frac{1}{2} \norm{ \bm{y}_{\txb} - \bm{A} \hat{\bm{x}}_{\txb} }_{2}^{2} 
        }
        = 
        F_{\txb} \paren{\alpha_{\txb}^{\ast}, \beta_{\txb}^{\ast}}, \label{eq:convergence_objective_binary} \\
        &\plim_{N \to \infty}\ 
        \frac{1}{N} \norm{\bm{y}_{\txb} - \bm{A} \hat{\bm{x}}_{\txb}}_{2}^{2} 
        = 
        (\beta_{\txb}^{\ast})^{2}, \label{eq:convergence_residual_binary}
    \end{align}
    respectively. 
\end{corollary}

Thus, we can estimate the noise variance in the binary vector reconstruction by using the proposed ARM. 
It should be noted that we do not require the tuning of the regularization parameter in this case because the optimization problem in~\eqref{eq:box_relaxation} does not contain any regularization parameter. 
Hence, the estimate of the noise variance can be obtained by the non-iterative approach as shown in Algorithm~\ref{alg:proposed_binary}. 
Note that the information of the noise variance is useful for other reconstruction methods such as~\cite{datta2010,datta2012,sasahara2017,hayakawa2017a}, though the box relaxation does not include any parameter to be tuned. 

\begin{algorithm}[t]
	\caption{Proposed ARM for binary vector reconstruction}
	\begin{algorithmic}[1]
		\Require{measurement vector $\bm{y}_{\txb}$, measurement matrix $\bm{A}$}
		\Ensure{estimated noise variance $\hat{\sigma}_{\txv}^{2}$}
        \State{Solve~\eqref{eq:box_relaxation} and obtain $\hat{\bm{x}}_{\txb}$.} 
        \State{${\displaystyle \Res(\hat{\bm{x}}_{\txb}) = \frac{1}{N} \norm{\bm{y}_{\txb} - \bm{A} \hat{\bm{x}}_{\txb}}_{2}^{2}}$}
        \State{${\displaystyle \hat{\sigma}_{\txv}^{2} = \argmin_{\sigma^{2} > 0}\ \abs{\beta_{\txb}^{\ast} \paren{\sigma^{2}}^{2} - \Res(\hat{\bm{x}}_{\txb}) }}$}
    \end{algorithmic}
	\label{alg:proposed_binary}
\end{algorithm}

The left hand sides of~\eqref{eq:convergence_objective_binary} and~\eqref{eq:convergence_residual_binary} differ only by a factor of two because $f_{\txb}(\hat{\bm{x}}_{\txb}) = 0$. 
It might be interesting to investigate the relation between the right-hand sides of~\eqref{eq:convergence_objective_binary} and~\eqref{eq:convergence_residual_binary}. 

The proposed ARM can also be applied to the binary sparse vector reconstruction and more general discrete-valued vector reconstruction~\cite{nagahara2015,atitallah2017,hayakawa2020a} by using Theorem~\ref{th:main} with the corresponding distribution $p_{X}$ and the proper regularizer $f(\cdot)$. 
%
%----------------
%  Simulation Results
%----------------
\section{Simulation Results} \label{sec:simulations}
In this section, we show some simulation results to demonstrate the performance of the proposed noise variance estimation. 
In the simulations, we compare the following methods. 
\begin{itemize}
    \item 
        \emph{ARM}: the noise variance estimation with the proposed ARM in Algorithms~\ref{alg:proposed} or~\ref{alg:proposed_binary}. 
        For the optimization of $\sigma^{2}$ in~\eqref{eq:optimization_sigma} and $\lambda$ in~\eqref{eq:lmd_prop},~\eqref{eq:lambda_update}, we use the solver \textsf{scipy.optimize.minimize\_scalar} for scalar minimization in scipy~\cite{2020SciPy-NMeth}. 
        The set $\Sigma$ in~\eqref{eq:lmd_prop} is fixed as $\Sigma = \curbra{10^{-5}, 10^{-3}, 10^{-1}}$ and $\lambda_{\mathrm{max}} = 1$ in the simulations. 
        The value of $\varepsilon$ in~\eqref{eq:diff} is set as $\varepsilon = 0.1$. 
    \item 
        \emph{AMP-LASSO}: the estimation method using AMP-LASSO given by~\eqref{eq:AMP-LASSO}~\cite{bayati2013}. 
        Since the tuning of the regularization parameter for AMP-LASSO has not been discussed in the paper, we use the proposed value $\lambda_{\prop}(\hat{p}_{0})$ unless otherwise stated. 
    \item 
        \emph{scaled residual}: the estimation method using the scaled residual given by~\eqref{eq:SR}. 
    \item 
        \emph{ridge regularization-based method}: the conventional SNR estimation method in~\cite{suliman2017} based on the asymptotic analysis of ridge regularized least squares. 
        The regularization parameters are the same as those in~\cite{suliman2017}. 
    \item 
        \emph{ML (oracle)}: the maximum likelihood (ML) approach when the true sparse vector $\bm{x}$ is known. 
        The estimate of $\sigma_{\txv}^{2}$ is given by $\hat{\sigma}_{\txv}^{2} = \frac{1}{M} \norm{\bm{y} - \bm{A}\bm{x}}_{2}^{2}$. 
        Note that $\bm{x}$ is unknown in the other methods. 
\end{itemize}
In all methods, the noise variance is estimated in the range $[10^{-6}, 1]$. 
The measurement matrix $\bm{A}$ and the noise vector $\bm{v}$ satisfy Assumption~\ref{ass:problem} in the simulations. 
\subsection{Sparse Vector Reconstruction}
\begin{figure*}[t]
    \begin{minipage}{0.33\textwidth}
        \centering
        \includegraphics[width=50mm]{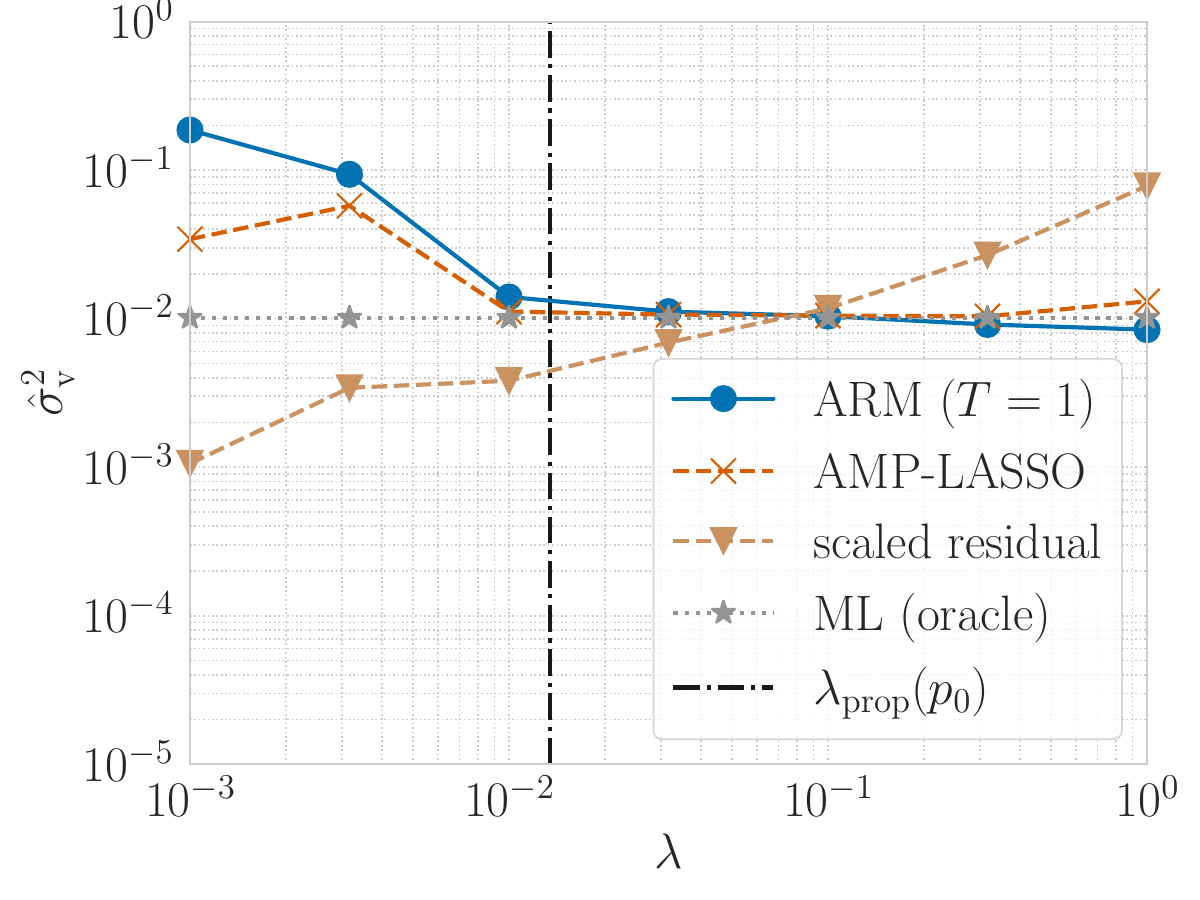}
        \subcaption{$\sigma_{\txv}^{2}=0.01$}
        \label{subfig:estimate_vs_lambda1}
    \end{minipage}
    \begin{minipage}{0.33\textwidth}
        \centering
        \includegraphics[width=50mm]{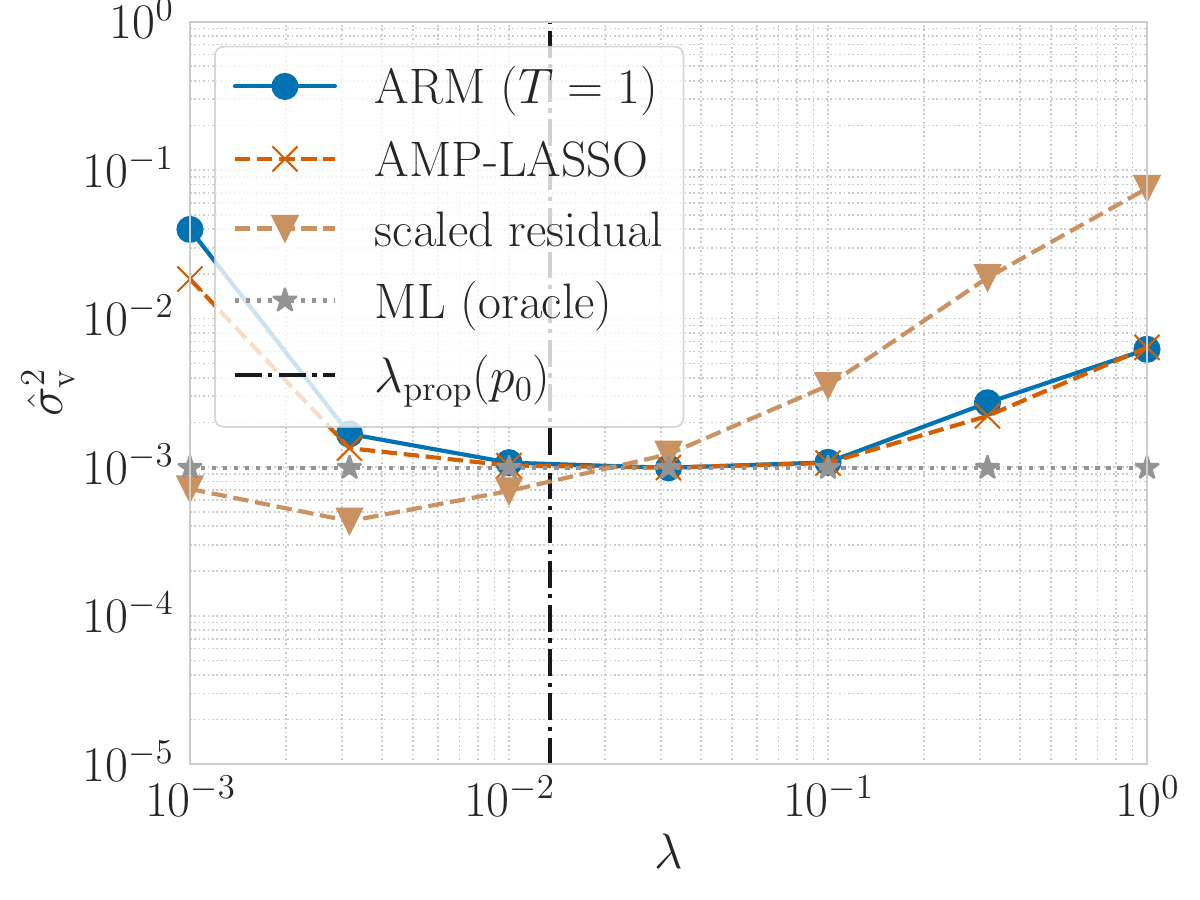}
        \subcaption{$\sigma_{\txv}^{2}=0.001$}
        \label{subfig:estimate_vs_lambda2}
    \end{minipage}
    \begin{minipage}{0.33\textwidth}
        \centering
        \includegraphics[width=50mm]{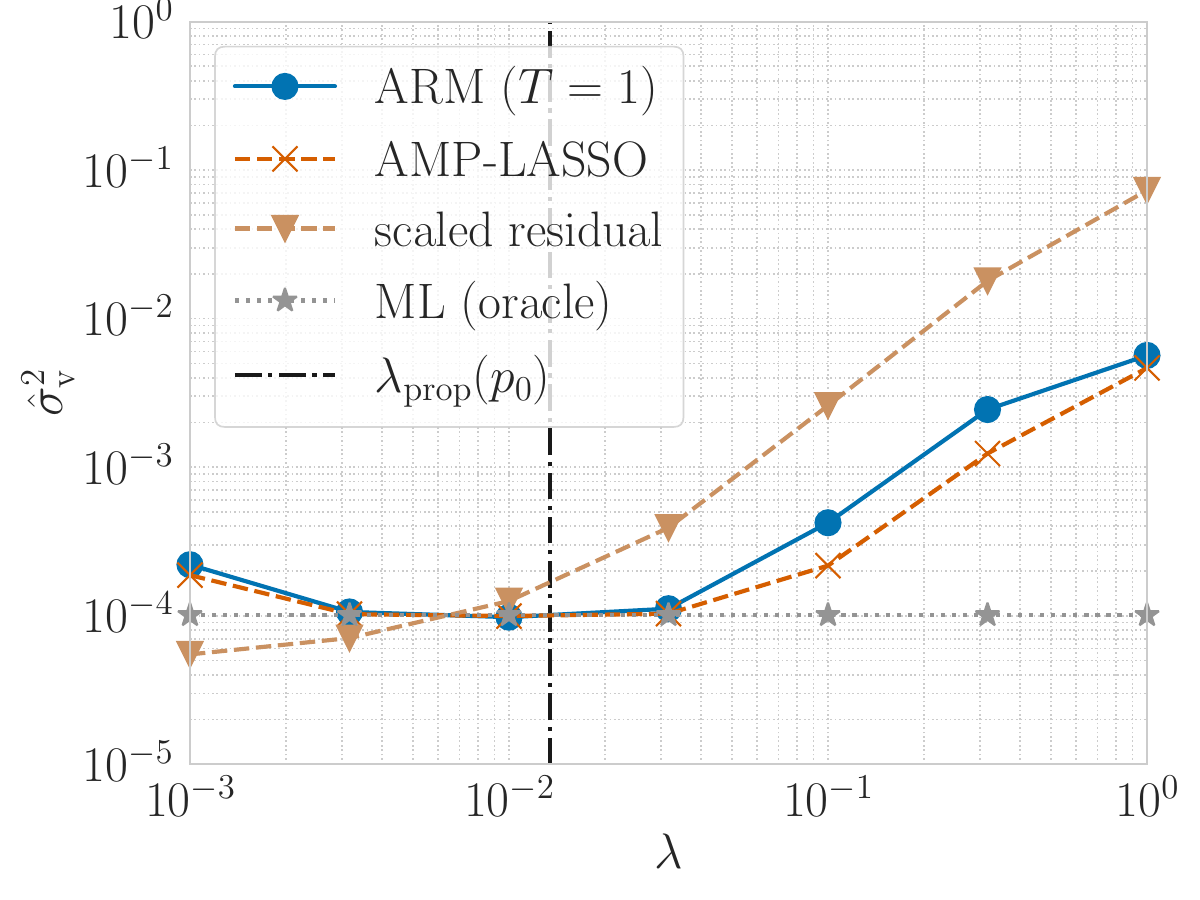}
        \subcaption{$\sigma_{\txv}^{2}=0.0001$}
        \label{subfig:estimate_vs_lambda3}
    \end{minipage}
    \caption{Estimated $\hat{\sigma}_{\txv}^{2}$ versus $\lambda$ ($N=200, \Delta=0.7, p_{0}=0.9$, $p_{\txX}(x)$: Bernoulli-Gaussian distribution).}
    \label{fig:estimate_vs_lambda}
\end{figure*}

We first examine the effect of the regularization parameter $\lambda$ in the noise variance estimation for the sparse vector reconstruction. 
Figure~\ref{fig:estimate_vs_lambda} shows the estimate of the noise variance $\hat{\sigma}_{\txv}^{2}$ versus the regularization parameter $\lambda$ for $N = 200$ and $\Delta = 0.7$. 
\begin{figure*}[t]
    \begin{minipage}{0.5\textwidth}
        \centering
        \includegraphics[width=80mm]{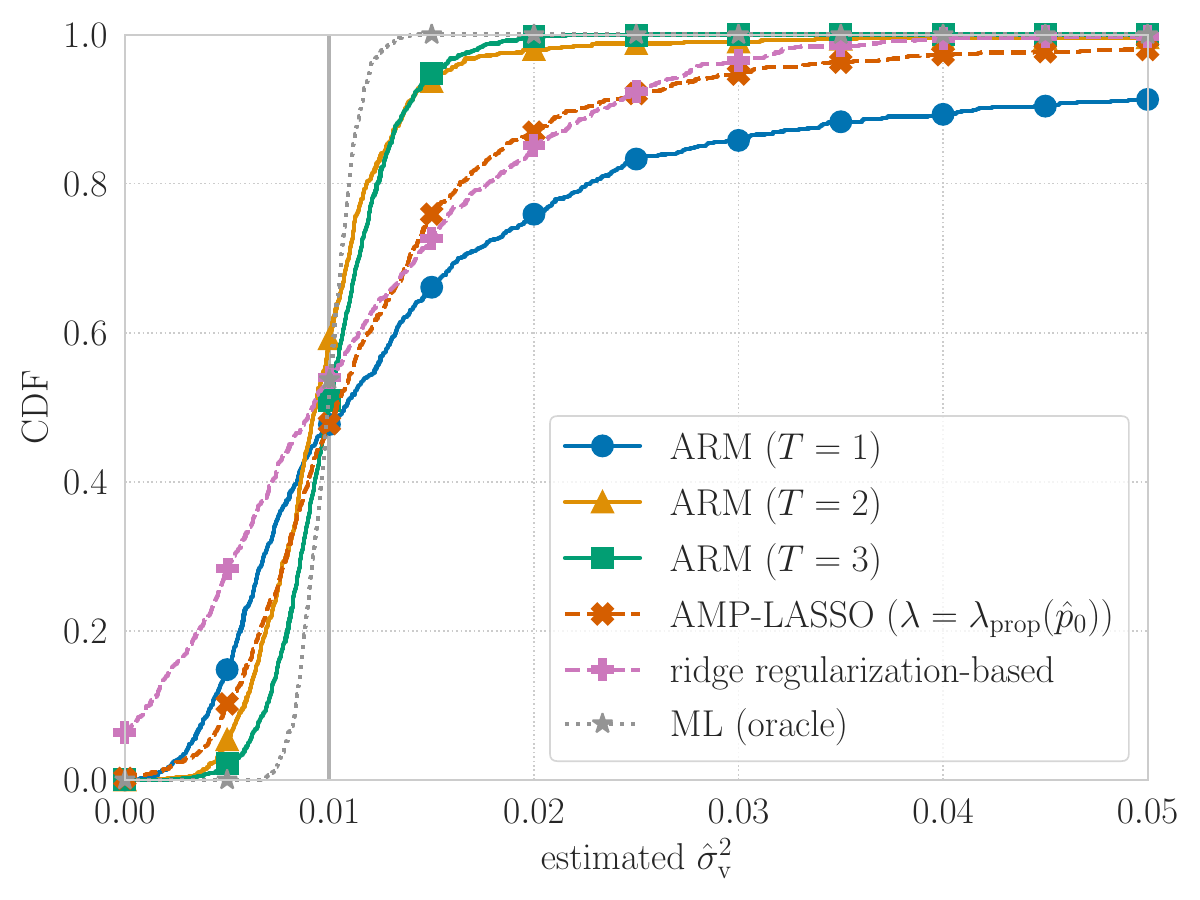}
        \subcaption{$\sigma_{\txv}^{2}=0.01$}
        \label{subfig:CDF1}
    \end{minipage}
    \begin{minipage}{0.5\textwidth}
        \centering
        \includegraphics[width=80mm]{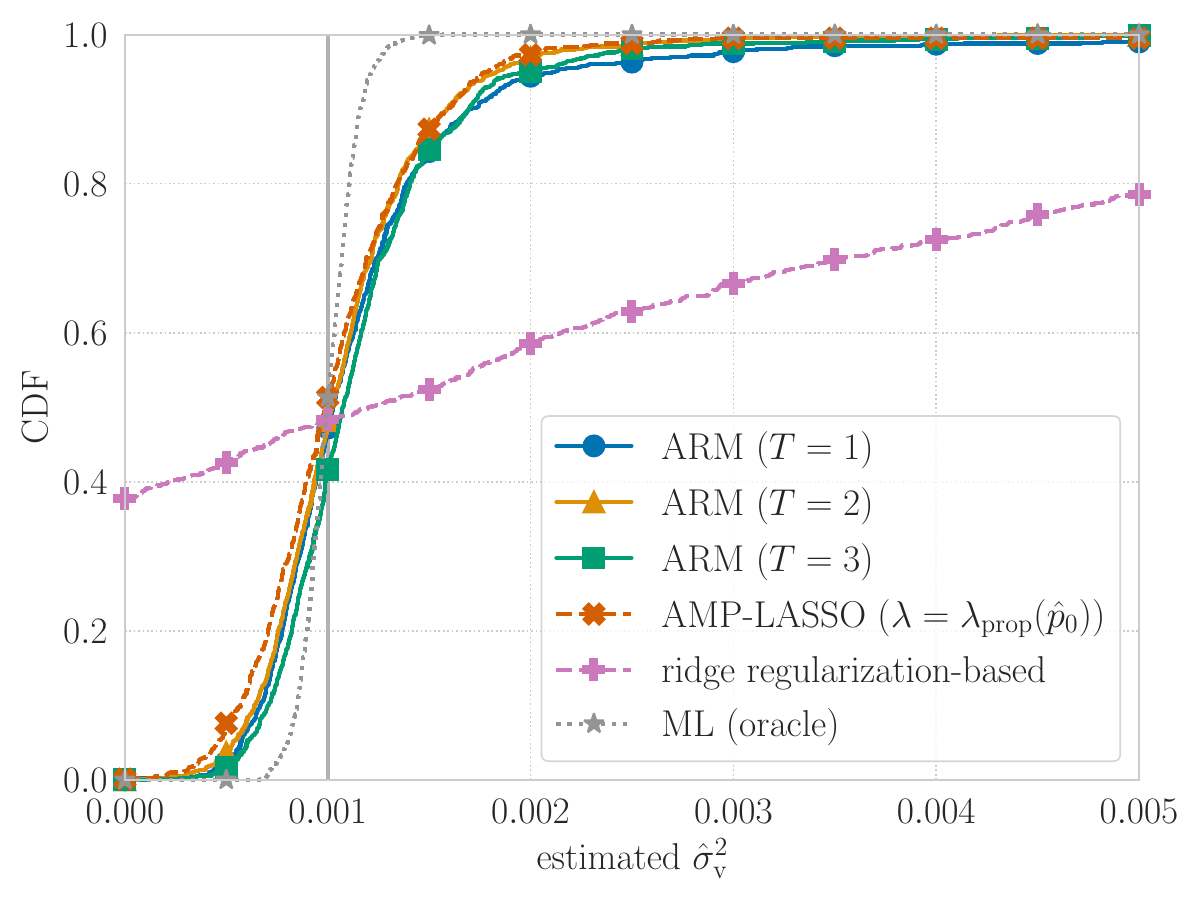}
        \subcaption{$\sigma_{\txv}^{2}=0.001$}
        \label{subfig:CDF2}
    \end{minipage}
    \caption{CDF of estimated $\hat{\sigma}_{\txv}^{2}$ ($N = 200$, $\Delta = 0.6$, $p_{0} = 0.9$, $p_{\txX}(x)$: Bernoulli-Gaussian distribution).}
    \label{fig:CDF}
\end{figure*}
The distribution of the unknown vector is the Bernoulli-Gaussian distribution in~\eqref{eq:Bernoulli-Gaussian} with $p_{0} = 0.9$. 
The true noise variance is set as $\sigma_{\txv}^{2} = 0.01, 0.001$, and $0.0001$ in Figs.~\ref{subfig:estimate_vs_lambda1},~\ref{subfig:estimate_vs_lambda2} and~\ref{subfig:estimate_vs_lambda3}, respectively. 
To solve~\eqref{eq:optimization} in ARM, AMP-LASSO, and scaled residual, we use the LASSO solver of scikit-learn~\cite{pedregosa2011scikit}. 
The estimated value is averaged over $100$ independent trials. 
In the figures, the black vertical line shows the value of the proposed regularization parameter $\lambda_{\prop}(p_{0})$ with the true probability $p_{0}$. 
Although the estimation performance depends on $\lambda$, the proposed regularization parameter can achieve good performance in both ARM and AMP-LASSO for $\sigma_{\txv}^{2} = 0.01, 0.001$, and $0.0001$. 
We can also see that the performance of the scaled residual is worse than the other methods. 

Figure~\ref{fig:CDF} shows the histogram of the empirical CDF of the estimated $\hat{\sigma}_{\txv}^{2}$, where $N =200$, $\Delta = 0.6$, and $p_{0} = 0.9$. 
The histogram is obtained from $1000$ independent trials. 
Since the true noise variance is set to $\sigma_{\txv}^{2} = 0.01$ in Fig.~\ref{subfig:CDF1}, it is better that the CDF rapidly increases around $\sigma_{\txv}^{2} = 0.01$. 
From the figure, we can see that the CDF of the proposed ARM with $T = 3$ increases around $\sigma_{\txv}^{2} = 0.01$ more rapidly than AMP-LASSO and the ridge regularization-based method. 
This means that the proposed method obtains the estimate near the true value with a higher probability. 
The figure also shows that the performance of the proposed ARM improves as the number of iterations $T$ increases. 
Figure~\ref{subfig:CDF2} shows the performance for $\sigma_{\txv}^{2} = 0.001$, where the proposed ARM and AMP-LASSO achieves similar performance. 
However, it should be noted that we use the proposed regularization parameter $\lambda_{\prop}(\hat{p}_{0})$ for AMP-LASSO. 
The performance of AMP-LASSO degrades if we use an inappropriate parameter value as shown in Fig.~\ref{fig:estimate_vs_lambda}. 
We can see that the proposed ARM with $T = 2, 3$ achieves a similar performance for both $\sigma_{\txv}^{2} = 0.01$ and $\sigma_{\txv}^{2} = 0.001$, whereas the performance of AMP-LASSO and the ridge regularization-based method largely depends on the true value of $\sigma_{\txv}^{2}$. 

\begin{figure}[t]
    \centering
    \includegraphics[width=80mm]{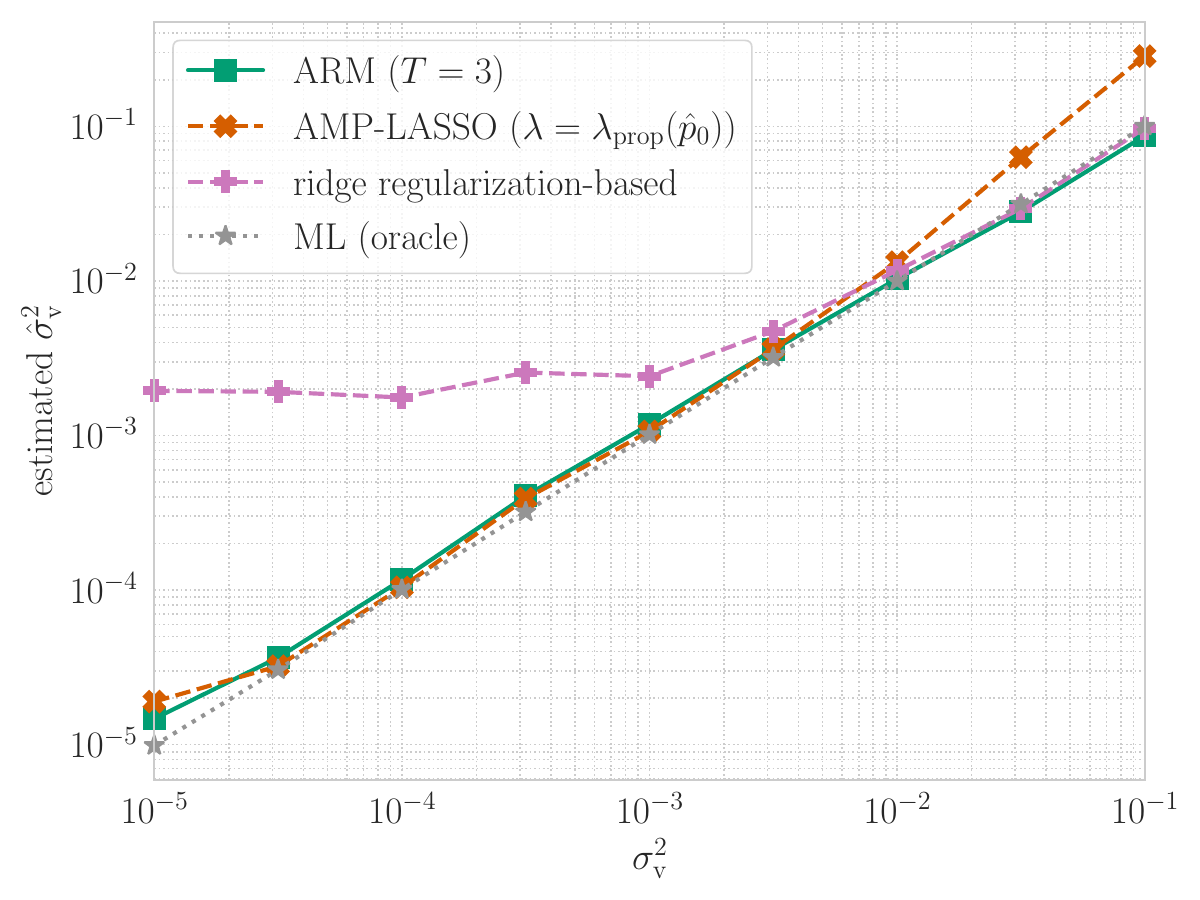}
    \caption{Estimated $\hat{\sigma}_{\txv}^{2}$ versus $\sigma_{\txv}^{2}$ ($N=200$, $\Delta=0.6$, $p_{0}=0.9$, $p_{\txX}(x)$: Bernoulli-Gaussian distribution).}
    \label{fig:estimate_vs_sigma2}
\end{figure}

We then evaluate the estimation performance for a wide range of noise variances. 
In Fig.~\ref{fig:estimate_vs_sigma2}, we plot the estimate $\hat{\sigma}_{\txv}^{2}$ versus its true value $\sigma_{\txv}^{2}$ when $N = 200$, $\Delta = 0.6$, and $p_{0} = 0.9$. 
The performance is obtained by averaging the result of $100$ independent trials. 
The figure shows that the proposed ARM with $T=3$ can achieve good estimation performance close to `ML (oracle)' for the whole range of $\sigma_{\txv}^{2}$ in the figure. 
On the other hand, the performance of AMP-LASSO and the ridge regularization-based method degrades for the large $\sigma_{\txv}^{2}$ and small $\sigma_{\txv}^{2}$, respectively. 

\begin{figure*}[t]
    \begin{minipage}{0.33\textwidth}
        \centering
        \includegraphics[width=50mm]{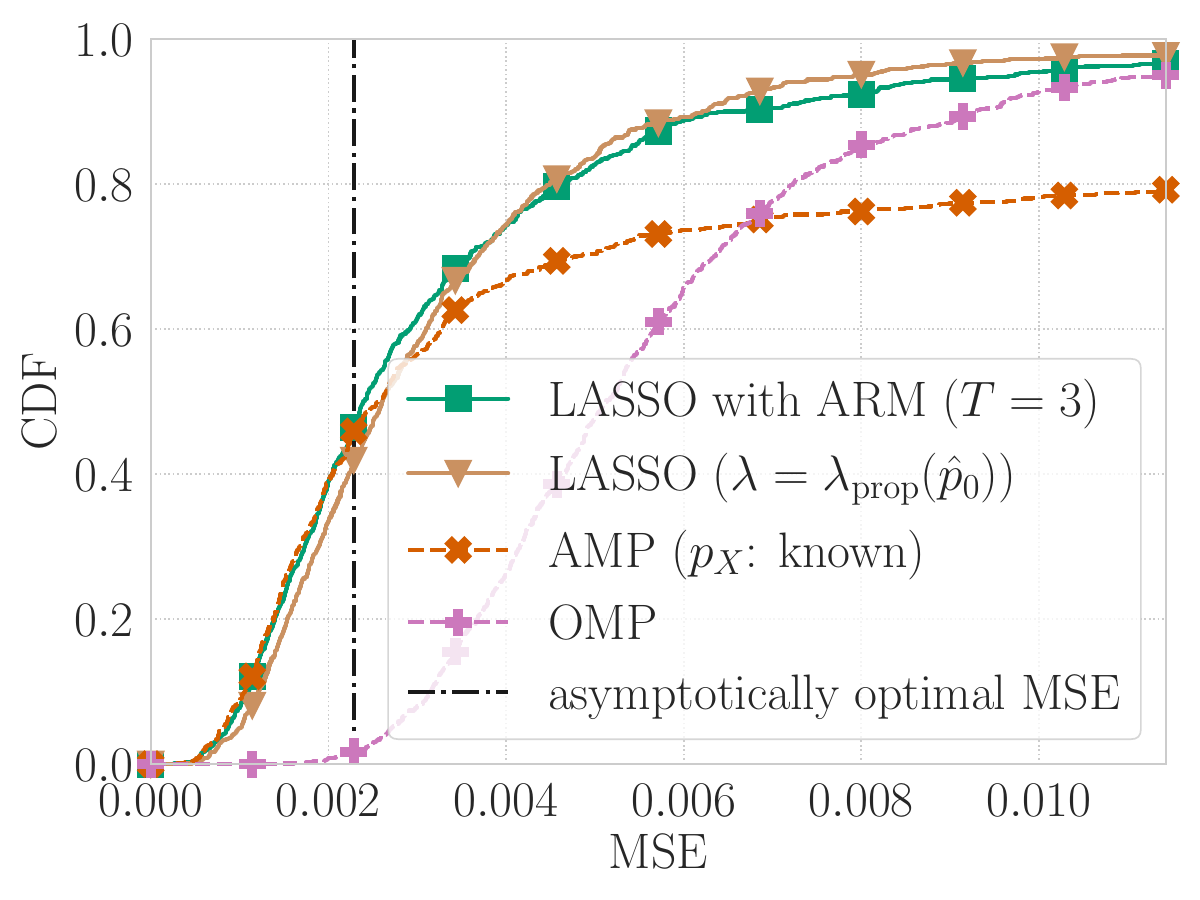}
        \subcaption{$N=100$}
        \label{subfig:MSE_CDF_001_1}
    \end{minipage}
    \begin{minipage}{0.33\textwidth}
        \centering
        \includegraphics[width=50mm]{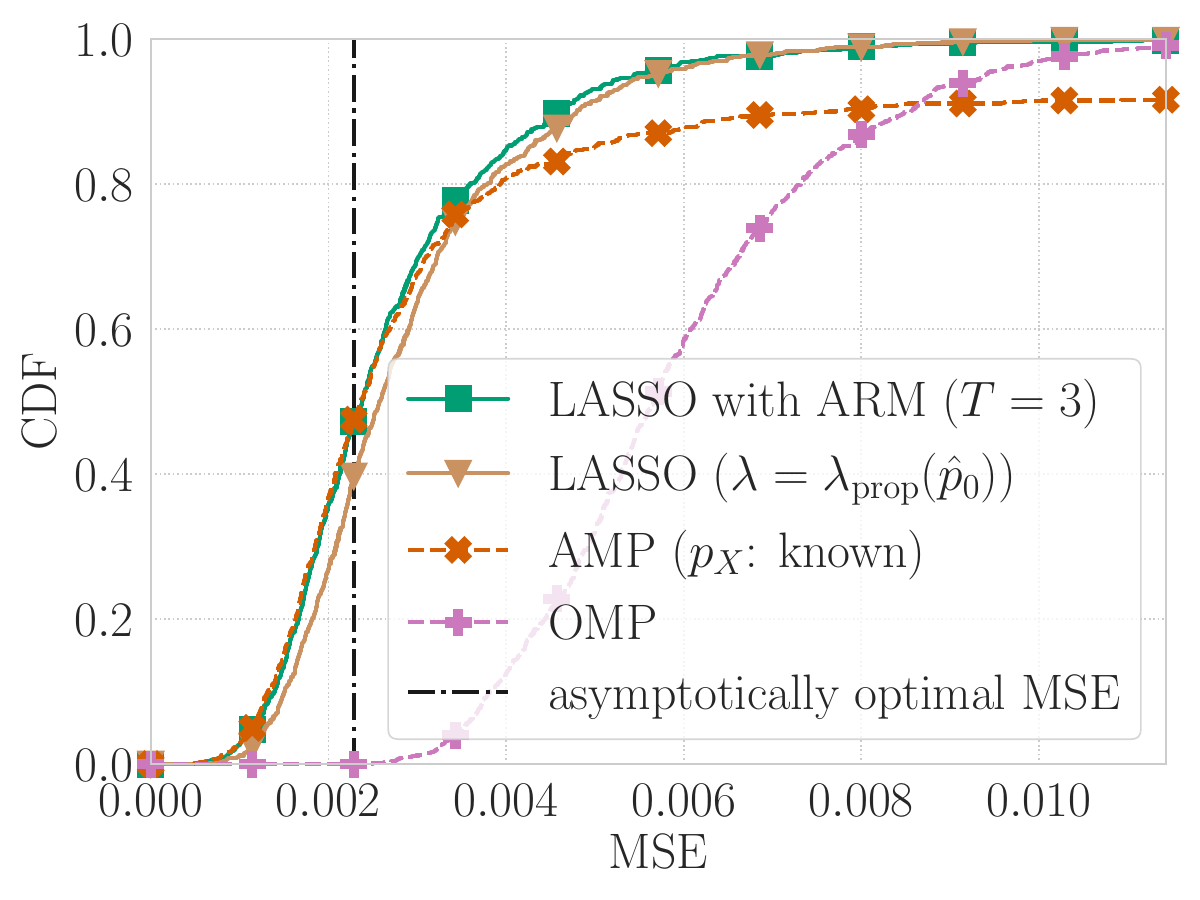}
        \subcaption{$N=200$}
        \label{subfig:MSE_CDF_001_2}
    \end{minipage}
    \begin{minipage}{0.33\textwidth}
        \centering
        \includegraphics[width=50mm]{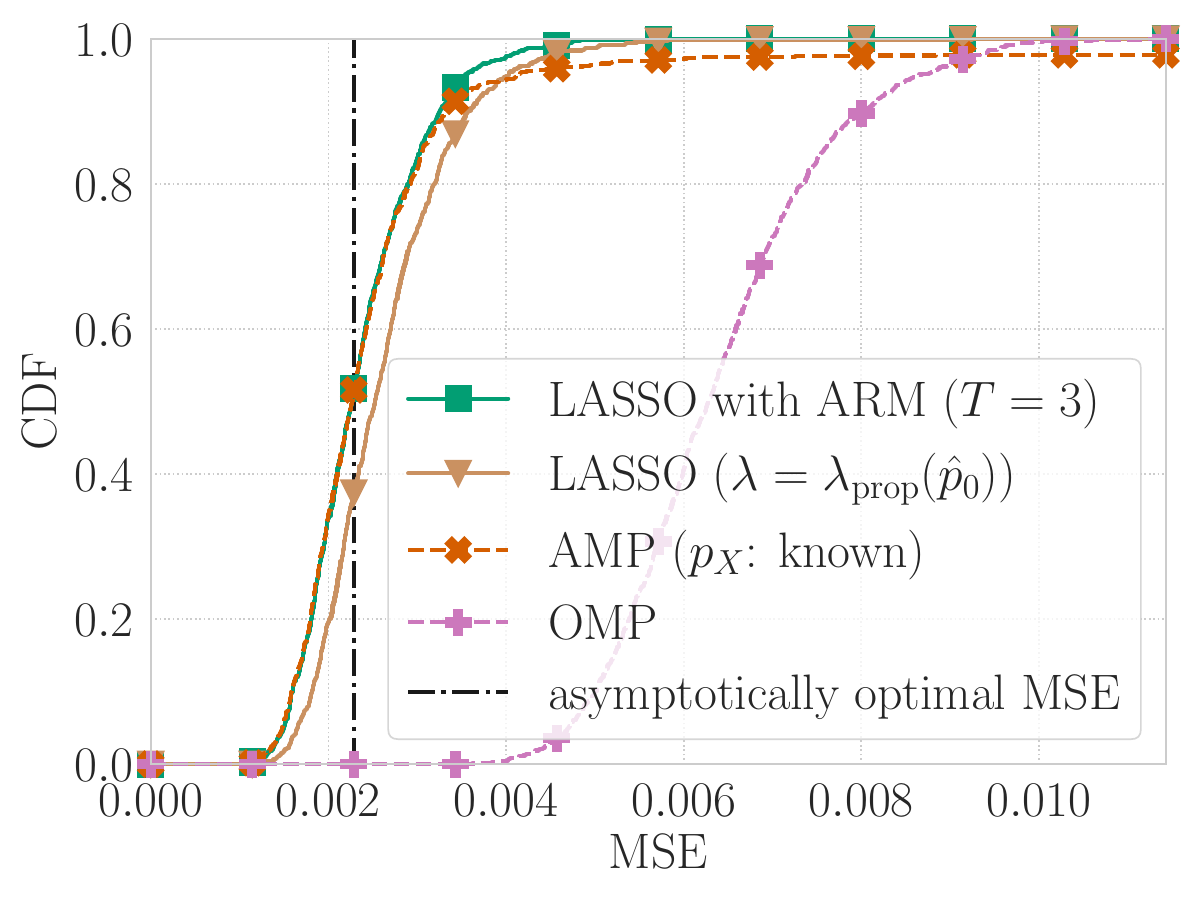}
        \subcaption{$N=500$}
        \label{subfig:MSE_CDF_001_3}
    \end{minipage}
    \caption{CDF of MSE ($\Delta=0.7$, $p_{0}=0.8$, $\sigma_{\txv}^{2}=0.001$, $p_{\txX}(x)$: Bernoulli-Gaussian distribution).}
    \label{fig:MSE_CDF_001}
\end{figure*}

Next, we demonstrate the reconstruction performance of the optimization problem in~\eqref{eq:optimization} with the proposed noise variance estimation. 
Figure~\ref{fig:MSE_CDF_001} shows the CDF of the MSE $\frac{1}{N} \norm{\hat{\bm{x}} - \bm{x}}_{2}^{2}$ ($\hat{\bm{x}}$: estimate of $\bm{x}$) obteind with $1000$ independent trials, where $\Delta = 0.7$, $p_{0} = 0.8$, and $\sigma_{\txv}^{2} = 0.001$. 
The dimension of the unknown vector is set as $N = 100$, $200$, and $500$ in Figs.~\ref{subfig:MSE_CDF_001_1},~\ref{subfig:MSE_CDF_001_2}, and~\ref{subfig:MSE_CDF_001_3}, respectively. 
In the figures, `LASSO with ARM' shows the performance of the optimization problem~\eqref{eq:optimization} with the parameter tuning by the proposed ARM. 
Specifically, we first obtain the estimate of the noise variance $\hat{\sigma}_{\txv}^{2}$ with the proposed ARM, and then calculate the optimal value of $\lambda$ in terms of asymptotic MSE by using the estimated $\hat{\sigma}_{\txv}^{2}$ and $\hat{p}_{0}$ via the CGMT framework. 
For comparison, we also plot the performance of LASSO with the proposed initial regularization parameter $\lambda = \lambda_{\mathrm{prop}}(\hat{p}_{0})$ in~\eqref{eq:lmd_prop} as `LASSO ($\lambda = \lambda_{\mathrm{prop}}(\hat{p}_{0})$)'. 
Moreover, we show the performance of the AMP algorithm with the optimal thresholding parameters~\cite{mousavi2018} as `AMP', for which the distribution of the unknown vector $p_{X}$ is assumed to be perfectly known. 
In addition, `OMP' denotes the performance of the OMP algorithm with the tolerance of $10^{-3}$, which is implemented by using the solver of scikit-learn. 
In the figure, the vertical black line shows the asymptotically optimal MSE, which can be obtained by the CGMT or AMP framework. 
From the figure, we can see that LASSO outperforms the other methods especially when $N$ is small. 
On the other hand, the CDF of the AMP algorithm is far from one when $N = 100$ and $N = 200$, which means that the AMP algorithm results in a large MSE or even diverges. 
This is because the large system limit is usually assumed in the AMP framework to obtain the low-complexity algorithm and the insightful analysis. 
Since the AMP algorithm achieves similar performance to LASSO when $N = 500$, it would be a suitable candidate for large-scale problems. 
The performance of the OMP algorithm is worse than the other methods, and hence somehow we need to choose an appropriate tolerance parameter. 
These results show that the proposed noise variance estimation enables us to obtain good reconstruction performance even when the true noise variance is unknown and the problem size is relatively small. 

\begin{figure*}[t]
    \begin{minipage}{0.33\textwidth}
        \centering
        \includegraphics[width=50mm]{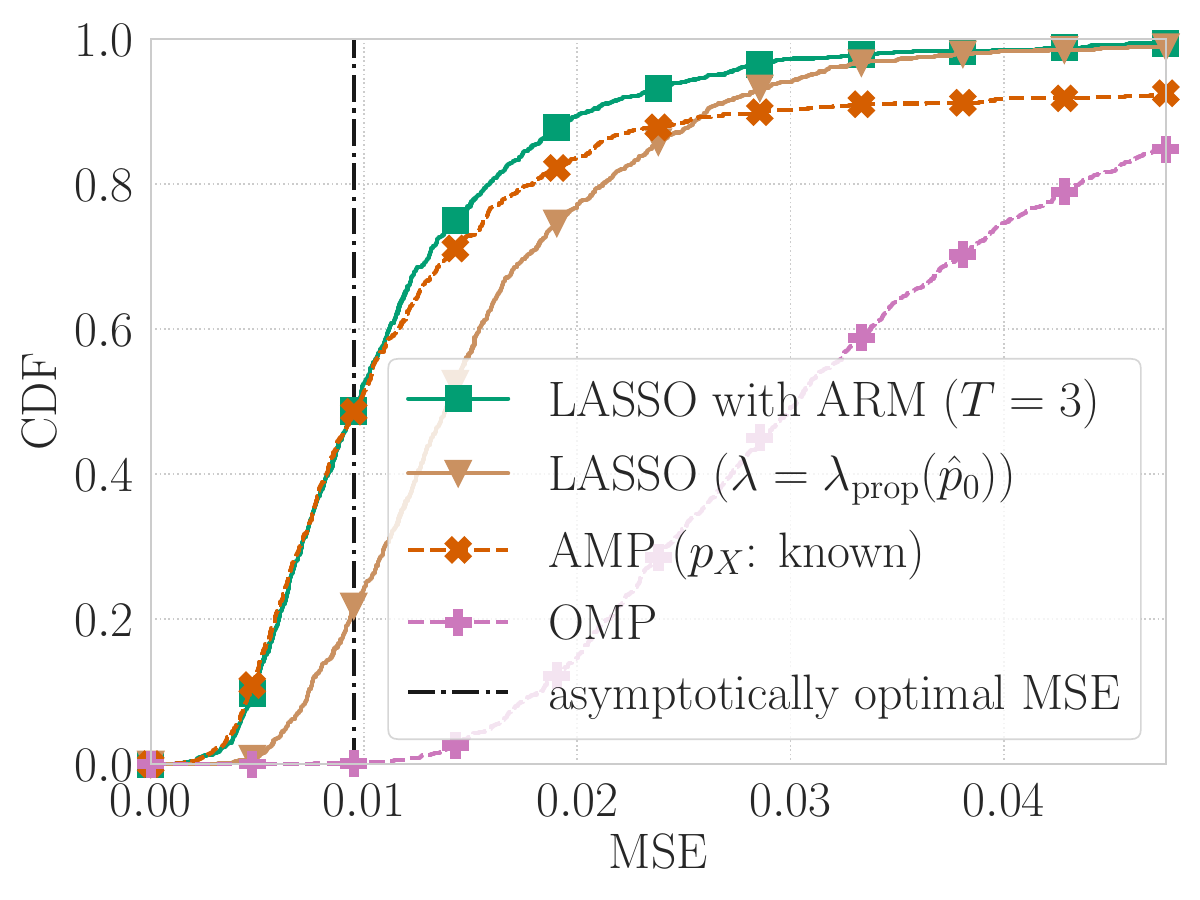}
        \subcaption{$N=100$}
        \label{subfig:MSE_CDF_005_1}
    \end{minipage}
    \begin{minipage}{0.33\textwidth}
        \centering
        \includegraphics[width=50mm]{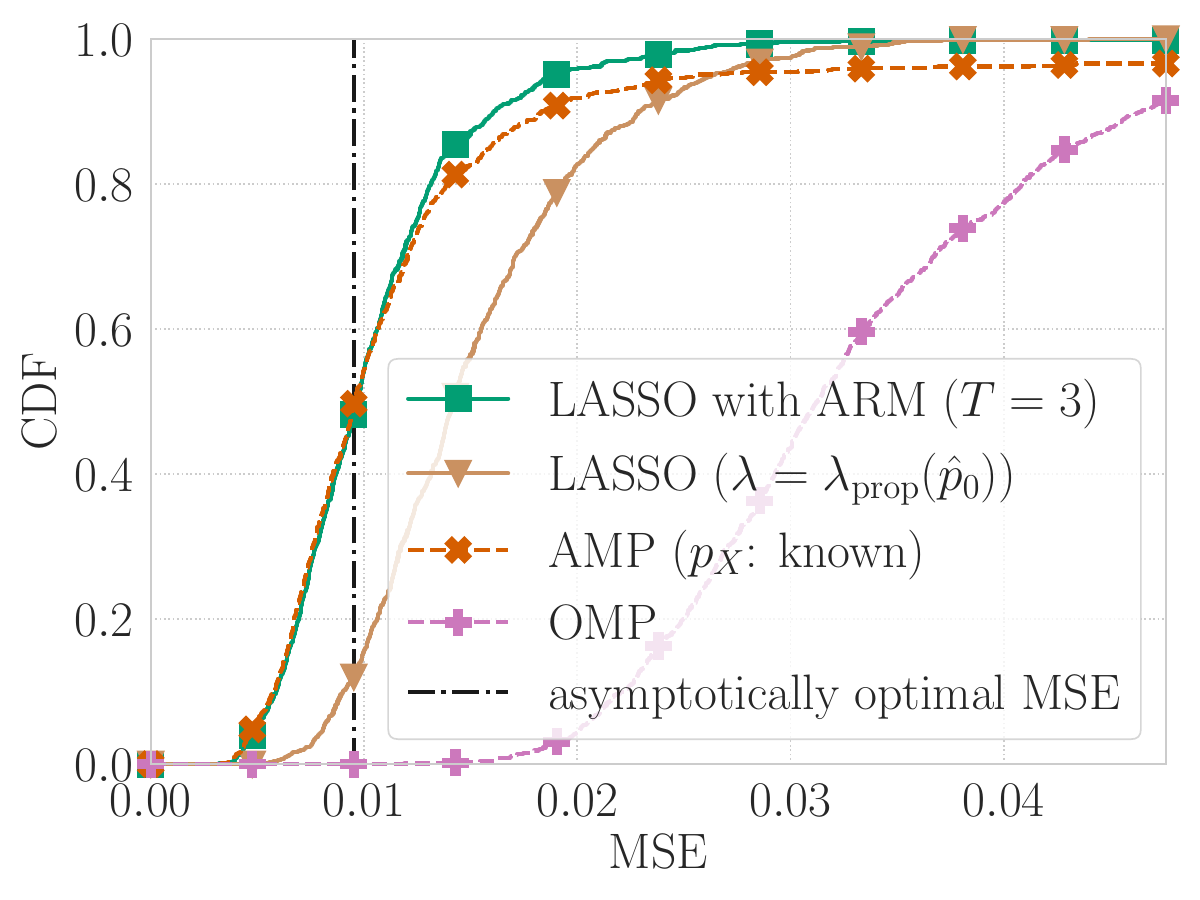}
        \subcaption{$N=200$}
        \label{subfig:MSE_CDF_005_2}
    \end{minipage}
    \begin{minipage}{0.33\textwidth}
        \centering
        \includegraphics[width=50mm]{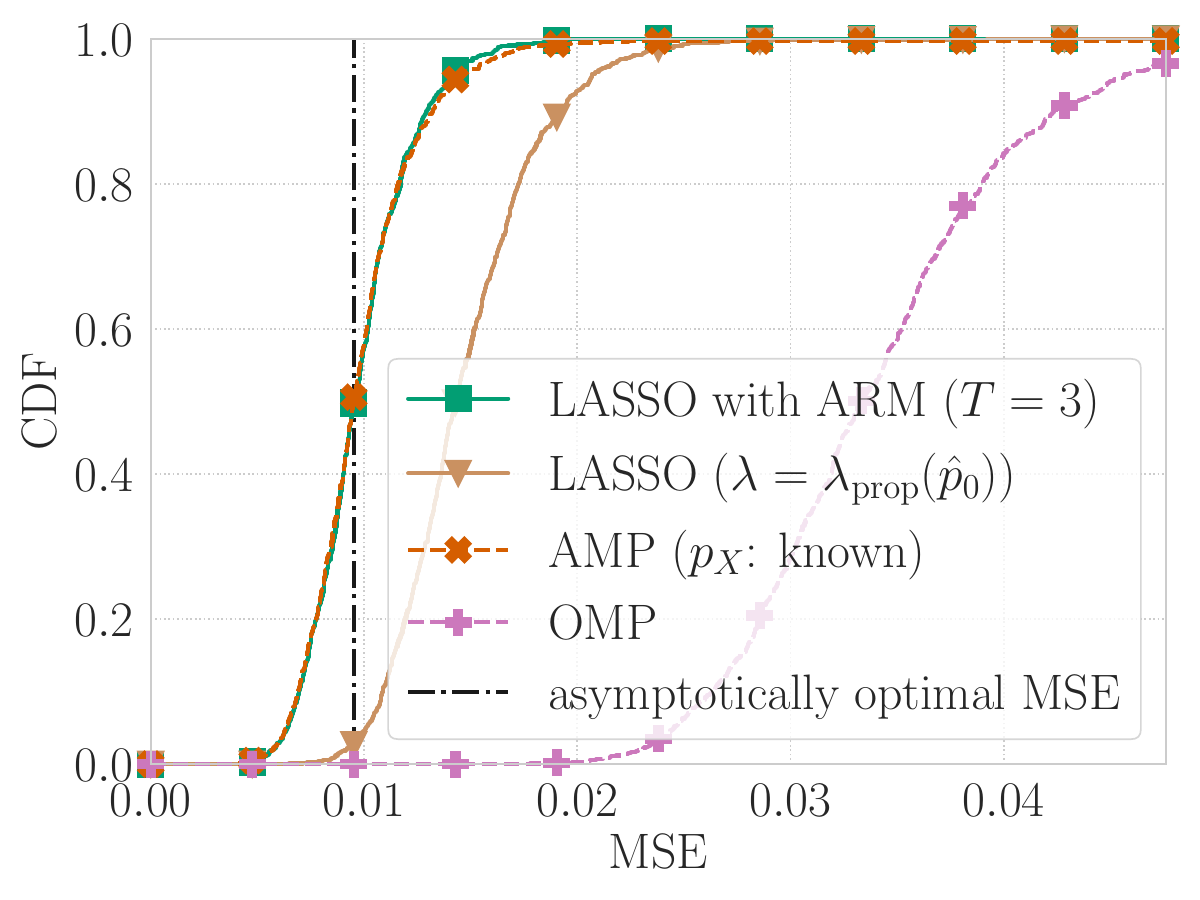}
        \subcaption{$N=500$}
        \label{subfig:MSE_CDF_005_3}
    \end{minipage}
    \caption{CDF of MSE ($\Delta=0.7$, $p_{0}=0.8$, $\sigma_{\txv}^{2}=0.005$, $p_{\txX}(x)$: Bernoulli-Gaussian distribution).}
    \label{fig:MSE_CDF_005}
\end{figure*}

Figure~\ref{fig:MSE_CDF_005} shows the CDF of the MSE obtained with $1000$ independent trials, where $\Delta = 0.7$, $p_{0} = 0.8$, and $\sigma_{\txv}^{2} = 0.005$. 
We obsereve that the performance of LASSO with $\lambda = \lambda_{\mathrm{prop}}(\hat{p}_{0})$ degrades compared to the case with Fig.~\ref{fig:MSE_CDF_001}. 
On the other hand, LASSO with ARM can achieve good performance even in this case, which shows the effectiveness of the noise variance estimation for the parameter tuning. 
\subsection{Binary Vector Reconstruction}
\begin{figure}[t]
    \centering
    \includegraphics[width=80mm]{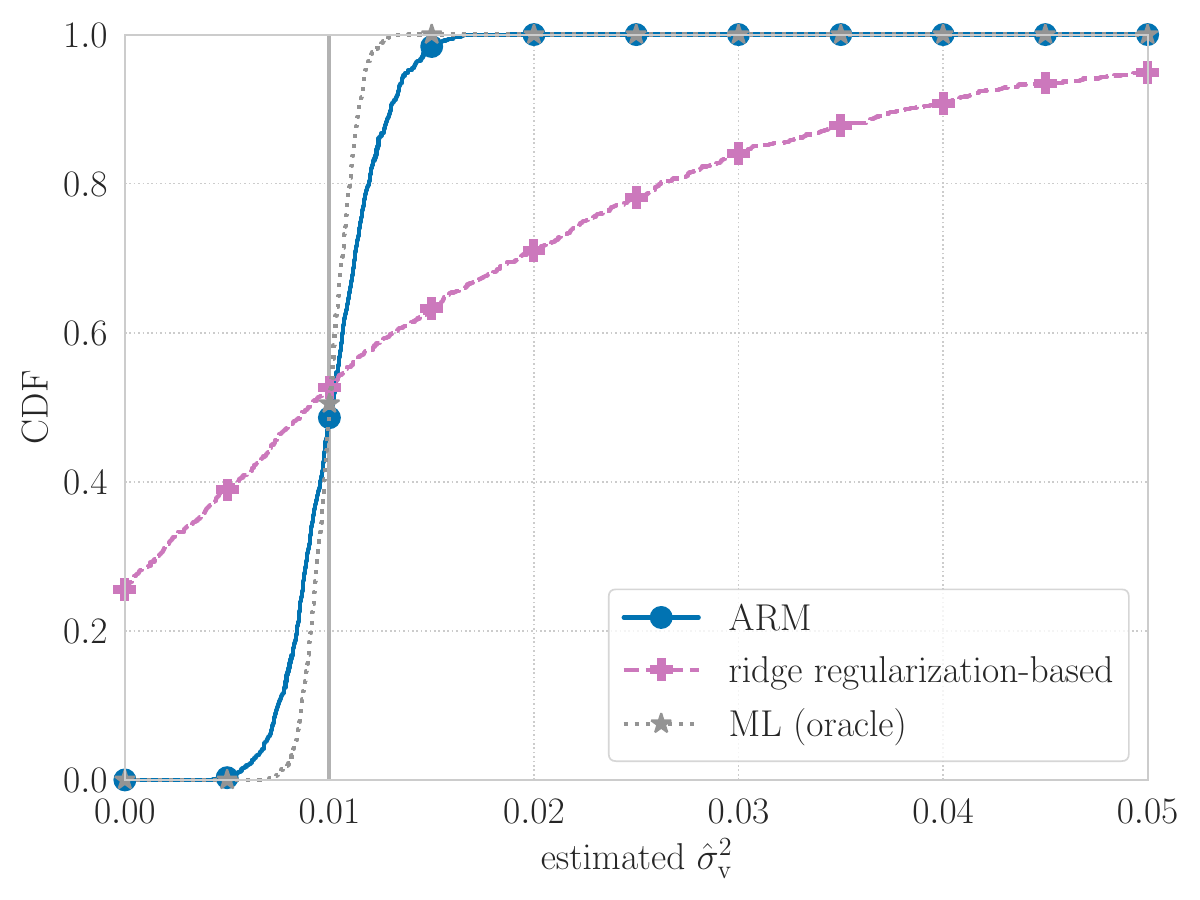}
    \caption{CDF of estimated $\hat{\sigma}_{\txv}^{2}$ ($N = 200$, $\Delta = 0.8$, $\sigma_{\txv}^{2}=0.01$, $p_{\txX, \txb}(x)$: Binary distribution).}
    \label{fig:CDF_binary}
\end{figure}

We then investigate the performance when the unknown vector is a binary vector with the distribution in~\eqref{eq:distribution_binary}. 
Figure~\ref{fig:CDF_binary} shows the histogram of the empirical CDF of the estimated $\hat{\sigma}_{\txv}^{2}$ when $N = 200$, $\Delta = 0.8$, and $\sigma_{\txv}^{2} = 0.01$. 
In the simulation, we use ADMM~\cite{gabay1976,eckstein1992,combettes2011,boyd2011} to solve the optimization problem~\eqref{eq:box_relaxation}. 
Since the estimate by AMP-LASSO in~\eqref{eq:AMP-LASSO} cannot be directly applied to the binary vector reconstruction, we compare the performance of the proposed method with the ridge-regularization based method~\cite{suliman2017}. 
As is the case with the Bernoulli-Gaussian distribution in Fig.~\ref{fig:CDF}, the proposed ARM in Algorithm~\ref{alg:proposed_binary} achieves better performance than the ridge regularization-based method. 

\begin{figure}[t]
    \centering
    \includegraphics[width=80mm]{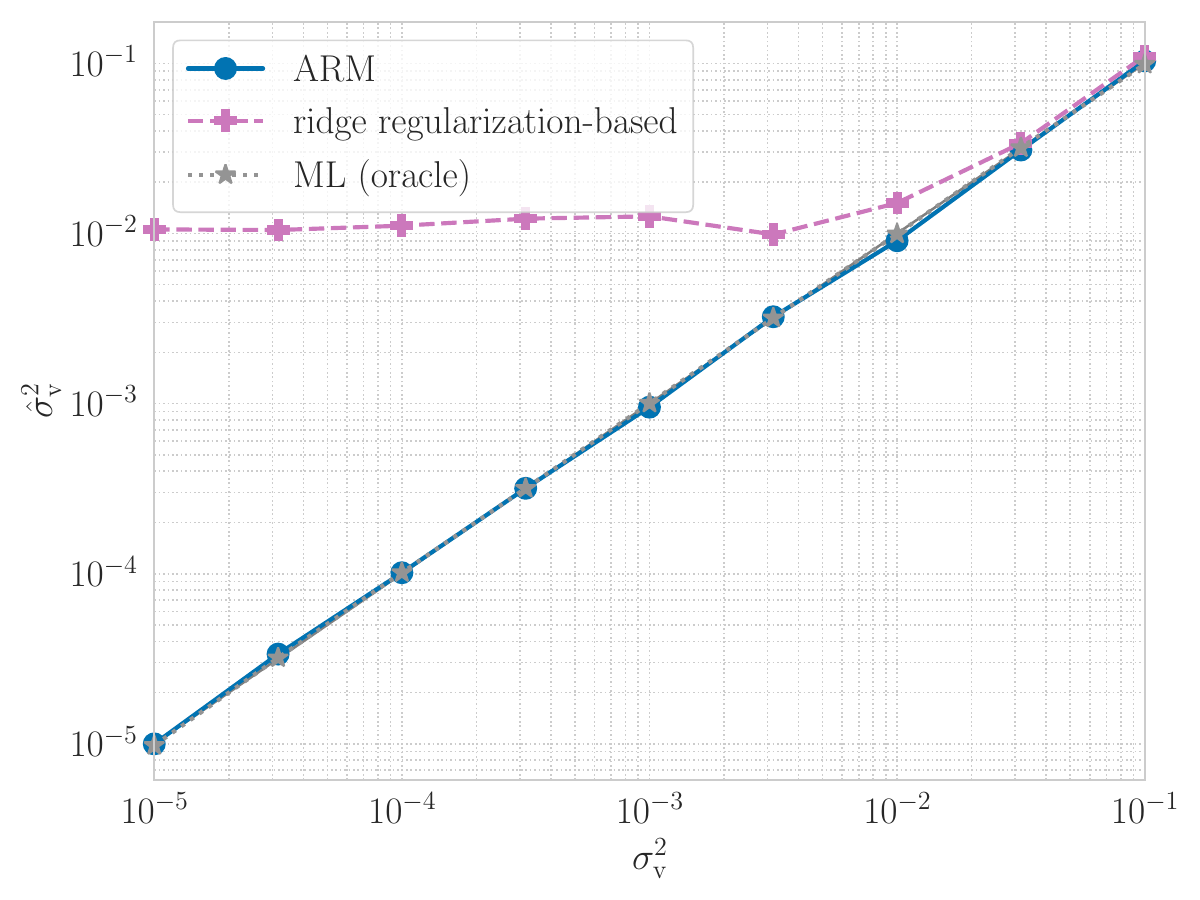}
    \caption{Estimated $\hat{\sigma}_{\txv}^{2}$ versus $\sigma_{\txv}^{2}$ ($N=200$, $\Delta=0.7$, $p_{\txX, \txb}(x)$: Binary distribution).}
    \label{fig:estimate_vs_sigma2_binary}
\end{figure}

Finally, we evaluate the estimation performance versus the true noise variance $\sigma_{\txv}^{2}$. 
Figure~\ref{fig:estimate_vs_sigma2_binary} shows the performance when $N = 200$ and $\Delta = 0.7$. 
The performance is obtained by averaging the result of $100$ independent trials. 
We observe that the proposed method achieves good estimation performance for a wide range of noise variances as is the case with Fig.~\ref{fig:estimate_vs_sigma2}. 
We thus conclude that the proposed noise variance estimation is effective for the binary distribution $p_{\txX, \txb}(x)$. 
%
%-------------
% Conclusion
%-------------
\section{Conclusion and Future Work} \label{sec:conclusion}
In this paper, we have proposed the noise variance estimation algorithm for compressed sensing with the Gaussian measurement matrix. 
The proposed ARM algorithm utilizes the asymptotic property of the estimate obtained by the $\ell_{1}$ optimization problem. 
Specifically, we estimate the noise variance by choosing the value whose corresponding asymptotic residual matches the empirical residual obtained by the actual reconstruction. 
The main advantages of the proposed approach can be summarized as follows: 
\begin{itemize}
    \item The proposed method can estimate a wide range of noise variances even in underdetermined problems. 
    \item We can design the choice of the regularization parameter $\lambda$ on the basis of the asymptotic results. 
    \item The proposed idea using the asymptotic residual can be extended for the reconstruction of some non-sparse structured vectors other than sparse ones as shown in Section~\ref{sec:extension}. 
    \item The proposed methods can achieve good performance even when the problem size is relatively small. 
\end{itemize}
Simulation results demonstrate that the proposed method can achieve better estimation performance than some conventional methods. 
Moreover, by using the estimate of the noise variance, we can choose an appropriate regularization parameter even when the noise variance is unknown. 
We have shown that the LASSO with the proposed noise variance estimation can achieve better performance than the AMP algorithm for small-scale problems. 

Compared to AMP-LASSO in~\eqref{eq:AMP-LASSO} and the scaled residual method in~\eqref{eq:SR}, the procedure of the proposed ARM is slightly complicated. 
For example, we need to estimate the probability $p_{0}$ of the unknown vector and solve some scalar optimization problems in the estimation. 
Although we have focused on the evaluation via computer simulations in this paper for the above reason, the proof of the consistency of the proposed method is important as a theoretical justification. 
Moreover, it would be an interesting research direction to apply the proposed idea for the choice of the regularization parameter $\lambda$ to the AMP-based methods. 
Although we have focused on the compressed sensing problem from the perspective of signal processing in this paper, the application of the proposed approach to statistics would also be a fascinating topic. 
The extension of the proposed method to the case with unknown distribution $p_{X}$ is also an important research direction. 
One possible approach is to iterate the estimation of $\sigma^{2}$ and $\bm{x}$ until the convergence, where we approximate the distribution $p_{X}$ with the empirical distribution of the estimated vector $\hat{\bm{x}}$. 

Extensions of the proposed approach to some variants of LASSO (e.g., constrained version) could be an interesting issue. 
It would be possible to apply the idea of the proposed approach to the case with other structured signals or other optimization problems because CGMT can be used for various optimization problems~\cite{thrampoulidis2015b,thrampoulidis2018,thrampoulidis2018a,thrampoulidis2018b,loureiro2022}. 
Since CGMT has also been applied to an optimization problem in the complex-valued domain~\cite{abbasi2019}, the extension to the complex-valued case could also be an interesting research direction. 
The generalization beyond the setting of Assumpsion~\ref{ass:problem}, e.g., partial Fourier measurements and non-i.i.d.\ measurement matrices~\cite{celentano2022,bellec2023}, would be also beneficial and left as an open problem. 
Application of conventional approaches as in~\cite{bellec2023} to binary vector reconstruction and the fair comparison with the proposed method would be important to reveal the advantage of each method. 
\appendices
% Proof of Theorem
\section{Proof of Theorem~\ref{th:main}} \label{app:proof}
In this section, we give the proof of Theorem~\ref{th:main}. 
Although the procedure of the proof partly follows some CGMT-based analyses (e.g., \cite{thrampoulidis2018, thrampoulidis2018a, hayakawa2020a}), we here show the sketch of the proof to derive the explicit formula in Theorem~\ref{th:main}. 
\subsection{CGMT}
We firstly summarize CGMT~\cite{thrampoulidis2015d,thrampoulidis2018} before the proof of Theorem~\ref{th:main}. 
CGMT associates the following primary optimization (PO) and auxiliary optimization (AO). 
\begin{align}
	(\text{PO}):\ & 
	\Phi(\bm{G}) 
	= 
	\min_{\bm{w} \in \mathcal{S}_{\txw}} \max_{\bm{u} \in \mathcal{S}_{\txu}} 
	\curbra{\bm{u}^{\top} \bm{G} \bm{w} + \xi(\bm{w},\bm{u})} \label{eq:PO_CGMT} \\
	(\text{AO}):\ & 
	\phi(\bm{g}, \bm{h}) 
	= 
	\min_{\bm{w} \in \mathcal{S}_{\txw}} \max_{\bm{u} \in \mathcal{S}_{\txu}} 
	\{ 
	    \norm{\bm{w}}_{2} \bm{g}^{\top} \bm{u} - \norm{\bm{u}}_{2} \bm{h}^{\top} \bm{w} \notag \\
        &\hspace{50mm}
	    + \xi(\bm{w}, \bm{u}) 
    \} \label{eq:AO_CGMT}
\end{align}
Here, $\bm{G} \in \mathbb{R}^{K \times L}$, $\bm{g} \in \mathbb{R}^{K}$, and $\bm{h} \in \mathbb{R}^{L}$ are composed of i.i.d.\ standard Gaussian variables. 
The constraint sets $\mathcal{S}_{\txw} \subset \mathbb{R}^{L}$ and $\mathcal{S}_{\txu} \subset \mathbb{R}^{K}$ are assumed to be closed compact. 
The function $\xi (\cdot,\cdot)$ is a continuous convex-concave function on $\mathcal{S}_{\txw} \times \mathcal{S}_{\txu}$. 

As in the following theorem, we can relate the optimal costs $\Phi(\bm{G})$, $\phi(\bm{g}, \bm{h})$ and the optimizer $\hat{\bm{w}}_{\Phi}(\bm{G})$ of (PO) (For more details, see~\cite[Theorem 3]{thrampoulidis2018} and~\cite[Theorem IV.2]{thrampoulidis2018a}). 
Intuitively, the theorem enables us to analyze (AO) instead of (AO). 
\begin{theorem}[CGMT] \label{th:CGMT} \mbox{}
    \begin{enumerate}
        \item 
            For all $\mu \in \mathbb{R}$ and $c > 0$, we have 
            \begin{align}
                \Pr (\abs{\Phi(\bm{G}) - \mu} > c) 
                \le 
                2 \Pr (\abs{\phi(\bm{g}, \bm{h}) - \mu} \ge c). \label{eq:CGMT_1}
            \end{align}
        \item 
            Let $\calS$ be a open set in $\calS_{\txw}$ and $\calS^{\txc} = \calS_{\txw} \setminus \calS$. 
            Moreover, we denote the optimal cost of (AO) with the constraint $\bm{w} \in \calS^{\txc}$ by $\phi_{\calS^{\txc}}(\bm{g}, \bm{h})$. 
            If there exists constants $\bar{\phi}$ and $\eta$ ($> 0$) such that $\phi(\bm{g}, \bm{h}) \le \bar{\phi} + \eta$ and $\phi_{\calS^{\txc}}(\bm{g}, \bm{h}) \ge \bar{\phi} + 2 \eta$ with probability approaching $1$ as $L \to \infty$, we then have 
            \begin{align}
                \lim_{L \to \infty} 
                \Pr (\hat{\bm{w}}_{\Phi}(\bm{G}) \in \calS) 
                = 
                1, \label{eq:CGMT_2}
            \end{align}
            where $L \to \infty$ means that $K$ and $L$ go to infinity with a fixed ratio. 
    \end{enumerate}
\end{theorem}
\subsection{(PO) Problem}
To obtain the result of Theorem~\ref{th:main} by using CGMT, we rewrite the $\ell_{1}$ optimization problem~\eqref{eq:optimization} as (PO) problem. 
We firstly define the error vector $\bm{u} = \bm{s} - \bm{x}$ and rewrite~\eqref{eq:optimization} as 
\begin{align}
    &\Phi_{N}^{\ast}
    \coloneqq 
    \min_{\bm{u} \in \mathbb{R}^{N}} 
    \frac{1}{N}
    \curbra{
        \frac{1}{2} \norm{ \bm{A}\bm{u} - \bm{v}}_{2}^{2} 
        + \lambda f (\bm{x} + \bm{u}) 
    }, \label{eq:optimization_error}
\end{align}
where the objective function is normalized by $N$. 
From \cite[Lemma~5]{thrampoulidis2018}, we can introduce a compact set $\calS_{\txu} \coloneqq \curbra{ \bm{u} \mid \norm{\bm{u}}_{2} \le C_{\txu} }$ with a constant $C_{\txu}$ ($> 0$) as 
\begin{align}
    &\Phi_{N}^{\ast}
    = 
    \min_{\bm{u} \in \calS_{\txu}} 
    \frac{1}{N}
    \curbra{
        \frac{1}{2} \norm{ \bm{A}\bm{u} - \bm{v}}_{2}^{2} 
        + \lambda f (\bm{x} + \bm{u}) 
    }. \label{eq:optimization_bounded}
\end{align}
Since we have 
\begin{align}
    \frac{1}{2} \norm{\bm{A} \bm{u} - \bm{v}}_{2}^{2} 
    &= 
    \max_{\bm{w} \in \mathbb{R}^{M}} 
    \curbra{
        \sqrt{N} \bm{w}^{\top} (\bm{A} \bm{u} - \bm{v}) 
        - \frac{N}{2} \norm{\bm{w}}_{2}^{2} 
    }, \label{eq:intro_u}
\end{align}
the optimization problem can be represented as 
\begin{align}
    \Phi_{N}^{\ast}
    &= 
    \min_{\bm{u} \in  \calS_{\txu}} \max_{\bm{w} \in \mathbb{R}^{M}} 
    \Biggl\{
        \frac{1}{N} \bm{w}^{\top} \paren{\sqrt{N} \bm{A}} \bm{u} 
        - \frac{1}{\sqrt{N}} \bm{v}^{\top}\bm{w} 
        \notag \\
    &\hspace{40mm}
        - \frac{1}{2} \norm{\bm{w}}_{2}^{2} 
        + \frac{\lambda}{N} f(\bm{x} + \bm{u})
    \Biggr\}. 
\end{align}
Moreover, by using \cite[Lemma~6]{thrampoulidis2018}, we can introduce a sufficiently large constraint set $\calS_{\txw} \coloneqq \curbra{ \bm{w} \mid \norm{\bm{w}}_{2} \le C_{\txw} }\ (C_{\txw} > 0)$ which will not affect the optimization problem with high probability as 
\begin{align}
    \Phi_{N}^{\ast}
    &= 
    \min_{\bm{u} \in  \calS_{\txu}} \max_{\bm{w} \in \calS_{\txw}} 
    \Biggl\{
        \frac{1}{N} \bm{w}^{\top} \paren{\sqrt{N} \bm{A}} \bm{u} 
        - \frac{1}{\sqrt{N}} \bm{v}^{\top}\bm{w} \notag \\\
    &\hspace{40mm}
        - \frac{1}{2} \norm{\bm{w}}_{2}^{2} 
        + \frac{\lambda}{N} f(\bm{x} + \bm{u})
    \Biggr\}. 
\end{align}
In the standard analysis based on CGMT, the minimization problem for the error vector $\bm{u}$ is analyzed. 
In our proof, however, we analyze the optimal value of $\bm{w}$ to obtain the result for the residual. 
We thus exchange the order of min-max from the minimax theorem and change the sign of the objective function to obtain 
\begin{align}
    -\Phi_{N}^{\ast}
    &= 
    \min_{\bm{w} \in \calS_{\txw}} \max_{\bm{u} \in \calS_{\txu}} 
    \Biggl\{
        \frac{1}{N} \bm{w}^{\top} \paren{\sqrt{N} \bm{A}} \bm{u} 
        + \frac{1}{\sqrt{N}} \bm{v}^{\top}\bm{w} \notag \\\
    &\hspace{40mm}
        + \frac{1}{2} \norm{\bm{w}}_{2}^{2} 
        - \frac{\lambda}{N} f(\bm{x} + \bm{u})
    \Biggr\}, \label{eq:PO}
\end{align}
where we can keep the sign of the first term $\frac{1}{N} \bm{w}^{\top} (\sqrt{N} \bm{A}) \bm{u}$ because the distribution of the matrix $\bm{A}$ is zero mean Gaussian and sign independent. 
The optimization problem~\eqref{eq:PO} is the form of (PO) normalized by $N$. 
Note that the optimal value of $\bm{w}$ can be written as 
\begin{align}
    \hat{\bm{w}}_{N}^{(\text{PO})} 
    &= 
    \frac{1}{\sqrt{N}} \paren{\bm{A} \hat{\bm{u}}_{N}^{(\text{PO})}  - \bm{v}} \\
    &= 
    \frac{1}{\sqrt{N}} \paren{\bm{A} \hat{\bm{x}}(\lambda) - \bm{y}} \label{eq:w_by_u}
\end{align}
from~\eqref{eq:intro_u}, where $\hat{\bm{u}}_{N}^{(\text{PO})} = \hat{\bm{x}}(\lambda) - \bm{x}$ is the optimal value of $\bm{u}$ in (PO). 
\subsection{(AO) Problem}
We then analyze the corresponding (AO) problem. 
Since the procedure is similar to~\cite{hayakawa2020a}, we omit some details in the analysis. 
The (AO) problem corresponding to~\eqref{eq:PO} is given by 
\begin{align}
    -\phi_{N}^{\ast} 
    &\coloneqq 
    \min_{\bm{w} \in \calS_{\txw}} \max_{\bm{u} \in \calS_{\txu}} 
    \Biggl\{
        \frac{1}{N} (\norm{\bm{w}}_{2} \bm{g}^{\top} \bm{u} 
        - \norm{\bm{u}}_{2} \bm{h}^{\top} \bm{w}) \notag \\  
    &\hspace{15mm}
        + \frac{1}{\sqrt{N}} \bm{v}^{\top} \bm{w} 
        + \frac{1}{2} \norm{\bm{w}}_{2}^{2} 
        - \frac{\lambda}{N} f(\bm{x} + \bm{u})
    \Biggr\}. \label{eq:AO}
\end{align}
Since the objective function in~\eqref{eq:AO} is not convex-concave, the order of min-max cannot be exchanged in general. 
As described in~\cite[Appendix A]{thrampoulidis2018}, however, we can flip the order in the asymptotic setting because the corresponding (PO) satisfies the condition for the min-max theorem. 
Hence, we exchange the order of min-max without detailed explanations hereafter. 
By exchanging the order of min-max and changing the sign of the objective function, we obtain 
\begin{align}
    \phi_{N}^{\ast} 
    &= 
    \min_{\bm{u} \in \calS_{\txu}} \max_{\bm{w} \in \calS_{\txw}} 
    \Biggl\{
        - \frac{1}{N} (\norm{\bm{w}}_{2} \bm{g}^{\top} \bm{u} 
        - \norm{\bm{u}}_{2} \bm{h}^{\top} \bm{w}) \notag \\ 
    &\hspace{15mm}
        - \frac{1}{\sqrt{N}} \bm{v}^{\top} \bm{w} 
        - \frac{1}{2} \norm{\bm{w}}_{2}^{2} 
        + \frac{\lambda}{N} f(\bm{x} + \bm{u})
    \Biggr\}. \label{eq:AO_flip}
\end{align}
Taking advantage of the fact that both $\bm{h}$ and $\bm{v}$ are Gaussian, we can rewrite $\frac{\norm{\bm{u}}_{2}}{\sqrt{N}} \bm{h} - \bm{v}$ as $\sqrt{\frac{\norm{\bm{u}}_{2}^{2}}{N} + \sigma_{\txv}^{2}}\ \bm{h}$, where we use the slight abuse of notation $\bm{h}$ as i.i.d.\ standard Gaussian variables. 
Using this technique, we can set $\norm{\bm{w}}_{2} = \beta$ and obtain the equivalent optimization problem
\begin{align}
    \phi_{N}^{\ast} 
    &= 
    \min_{\bm{u} \in \calS_{\txu}} \max_{\beta \ge 0} 
    \Biggl\{
        \sqrt{\frac{\norm{\bm{u}}_{2}^{2}}{N} + \sigma_{\txv}^{2}} \frac{\beta \norm{\bm{h}}_{2}}{\sqrt{N}} 
        - \frac{1}{N} \beta \bm{g}^{\top} \bm{u} 
        \notag \\
    &\hspace{45mm}
        - \frac{1}{2} \beta^{2} 
        + \frac{\lambda}{N} f(\bm{x} + \bm{u})
    \Biggr\}. \label{eq:AO2}
\end{align}
To further rewrite the optimization problem~\eqref{eq:AO2}, we use the following identity 
\begin{align}
    \chi 
    &= 
    \min_{\alpha>0}
    \paren{
        \frac{\alpha}{2} + \frac{\chi^{2}}{2\alpha} \label{eq:square-root-trick}
    } 
\end{align}
for $\chi = \sqrt{\frac{\norm{\bm{u}}_{2}^{2}}{N} + \sigma_{\txv}^{2}}$ and obtain 
\begin{align}
    &\min_{\alpha > 0} \max_{\beta \ge 0} 
    \Biggl\{
        \frac{\alpha \beta}{2} \frac{\norm{\bm{h}}_{2}}{\sqrt{N}} 
        + \frac{\sigma_{\txv}^{2} \beta}{2 \alpha} \frac{\norm{\bm{h}}_{2}}{\sqrt{N}} 
        - \frac{1}{2} \beta^{2}  \notag \\
    &\hspace{5mm}
        - \frac{1}{N} \frac{\alpha \beta \norm{\bm{g}}_{2}^{2}}{2} \frac{\sqrt{N}}{\norm{\bm{h}}_{2}}
        + \frac{\beta}{\alpha} \frac{\norm{\bm{h}}_{2}}{\sqrt{N}} 
        \frac{1}{N} \sum_{n=1}^{N} 
        \min_{u_{n} \in \mathbb{R}} J_{n}(u_{n})
    \Biggr\}, \label{eq:AO3}
\end{align}
where we define 
\begin{align}
    J_{n}(u_{n}) 
    &= 
    \frac{1}{2} 
    \paren{u_{n} - \frac{\sqrt{N}}{\norm{\bm{h}}_{2}} \alpha g_{n}}^{2} 
    + 
    \frac{\alpha \lambda}{\beta} \frac{\sqrt{N}}{\norm{\bm{h}}_{2}} f(x_{n} + u_{n}). \label{eq:J}
\end{align}
Here, $u_{n}$ and $g_{n}$ are the $n$-th element of $\bm{u}$ and $\bm{g}$, respectively. 
Note that we have exchanged the order of min-max from~\eqref{eq:AO2} with~\eqref{eq:square-root-trick} to~\eqref{eq:AO3} by using the fact that the objective function is convex for $\alpha, \bm{u}$ and concave for $\beta$ (For a similar and detailed discussion, see~\cite[Eq. (57)]{hayakawa2022}). 
Since we have 
\begin{align}
    \min_{u_{n} \in \mathbb{R}} J_{n}(u_{n}) 
    &= 
    \env_{\frac{\alpha \lambda}{\beta}\frac{\sqrt{N}}{\norm{\bm{h}}_{2}}f} 
    \paren{x_{n} + \frac{\sqrt{N}}{\norm{\bm{h}}_{2}} \alpha g_{n}}, 
\end{align}
the (AO) problem can be written as 
\begin{align}
    \phi_{N}^{\ast} 
    &= 
    \min_{\alpha > 0} \max_{\beta \ge 0} F_{N}(\alpha, \beta), 
\end{align}
where 
\begin{align}
    &F_{N}(\alpha, \beta) \notag \\
    &= 
    \frac{\alpha \beta}{2} \frac{\norm{\bm{h}}_{2}}{\sqrt{N}} 
    + \frac{\sigma_{\txv}^{2} \beta}{2 \alpha} \frac{\norm{\bm{h}}_{2}}{\sqrt{N}} 
    - \frac{1}{2} \beta^{2} 
    - \frac{1}{N} \frac{\alpha \beta \norm{\bm{g}}_{2}^{2}}{2} \frac{\sqrt{N}}{\norm{\bm{h}}_{2}} \notag \\
    &\hspace{4mm} 
    + \frac{\beta}{\alpha} \frac{\norm{\bm{h}}_{2}}{\sqrt{N}} 
    \frac{1}{N} \sum_{n=1}^{N} 
    \env_{\frac{\alpha \lambda}{\beta}\frac{\sqrt{N}}{\norm{\bm{h}}_{2}}f} 
    \paren{x_{n} + \frac{\sqrt{N}}{\norm{\bm{h}}_{2}} \alpha g_{n}}. \label{eq:F_N}
\end{align}
We denote the optimal values of $\alpha$ and $\beta$ in the (AO) problem by $\alpha_{N}^{\ast}$ and $\beta_{N}^{\ast}$, respectively. 
\subsection{Applying CGMT}
By using the above analysis, we confirm~\eqref{eq:convergence_objective} in Theorem~\ref{th:main}. 
As $N \to \infty$, $F_{N}(\alpha, \beta)$ in~\eqref{eq:F_N} converges pointwise to $F(\alpha, \beta)$ in~\eqref{eq:SO}. 
Letting $\phi^{\ast} = F(\alpha^{\ast}, \beta^{\ast})$ be the optimal value of $F(\alpha, \beta)$, we can obtain $-\phi_{N}^{\ast} \Pto -\phi^{\ast}$ and $(\alpha_{N}^{\ast}, \beta_{N}^{\ast}) \Pto (\alpha^{\ast}, \beta^{\ast})$ as $N \to \infty$ by a similar approach to the proof of~\cite[Lemma IV. 1]{thrampoulidis2018a}. 
Hence, by setting $\mu = -\phi^{\ast}$ in~\eqref{eq:CGMT_1} of Theorem~\ref{th:CGMT}, we have $\lim_{N \to \infty} \Pr (\abs{-\Phi_{N}^{\ast} - (-\phi^{\ast})} > c) = 0$ for any $c > 0$, which means~\eqref{eq:convergence_objective}. 

We can also demonstrate the convergence of the residual in~\eqref{eq:convergence_residual} from the second statement in Theorem~\ref{th:CGMT}. 
We denote the optimal value of $\bm{w}$ in~\eqref{eq:AO} by $\hat{\bm{w}}_{N}^{(\text{AO})}$ and define 
\begin{align}
    \calS 
    &= 
    \curbra{
        \bm{z} \in \mathbb{R}^{M} \Bigg| 
        \abs{\norm{\bm{z}}_{2}^{2} - \paren{\beta^{\ast}}^{2}} < \varepsilon
    }. 
\end{align}
We then have $\hat{\bm{w}}_{N}^{(\text{AO})} \in \calS$ with probability approaching $1$ for any $\varepsilon$ ($> 0$) because $\norm{\hat{\bm{w}}_{N}^{(\text{AO})}}_{2} = \beta_{N}^{\ast}$ from the definition of $\beta$ and $\beta_{N}^{\ast} \Pto \beta^{\ast}$. 
Considering the strong concavity of the objective function in~\eqref{eq:AO2} over $\beta$, we can see that there exists $\eta$ ($> 0$) satisfying the condition in Theorem~\ref{th:CGMT} with $\bar{\phi} = -\phi^{\ast}$. 
We thus have $\lim_{N \to \infty} \Pr\paren{\hat{\bm{w}}_{N}^{(\text{PO})} \in \calS} = 1$, i.e., 
\begin{align}
    \plim_{N \to \infty}\ 
    \norm{\hat{\bm{w}}_{N}^{(\text{PO})}}_{2}^{2} 
    = 
    (\beta^{\ast})^{2}. \label{eq:convergence_w_PO}
\end{align}
Combining~\eqref{eq:w_by_u} and~\eqref{eq:convergence_w_PO} concludes the proof. 
\section{On Expectation in~\eqref{eq:SO}} \label{app:exp}
In this section, we derive the explicit formula of the expectation in~\eqref{eq:SO} for the Bernoulli-Gaussian distribution in~\eqref{eq:Bernoulli-Gaussian}. 
The expectation in~\eqref{eq:SO} can be written as 
\begin{align}
    &\Ex{\env_{\frac{\alpha \lambda}{\beta \sqrt{\Delta}} f} \paren{X + \frac{\alpha}{\sqrt{\Delta}} G}} \notag \\
    &= 
    \frac{\alpha \lambda}{\beta \sqrt{\Delta}} 
    \Ex{f \paren{\prox_{\frac{\alpha \lambda}{\beta \sqrt{\Delta}} f} \paren{X + \frac{\alpha}{\sqrt{\Delta}} G}}} \notag \\
    &\hspace{2mm}
    + 
    \frac{1}{2} \mathrm{E} \sqbra{ \paren{ \prox_{\frac{\alpha \lambda}{\beta \sqrt{\Delta}} f} \paren{X + \frac{\alpha}{\sqrt{\Delta}} G}  
    -
    \paren{X + \frac{\alpha}{\sqrt{\Delta}} G} }^{2} }. \label{eq:Ex}
\end{align}
Since the proximity operator of $\gamma f(\cdot)$ ($\gamma > 0$) is given by 
\begin{align}
    \prox_{\gamma f} (q) 
    &= 
    \sign (q) \max(\abs{q} - \gamma, 0), \label{eq:prox_L1}
\end{align}
the expectation in the first term of~\eqref{eq:Ex} can be further rewritten as 
\begin{align}
    &\Ex{f \paren{\prox_{\frac{\alpha \lambda}{\beta \sqrt{\Delta}} f} \paren{X + \frac{\alpha}{\sqrt{\Delta}} G}}} \notag \\
    &= 
    p_{0} \int_{-\infty}^{\infty} \abs{\prox_{\frac{\alpha \lambda}{\beta \sqrt{\Delta}} f} \paren{\frac{\alpha}{\sqrt{\Delta}} g}} p_{\txG}(g) dg \notag \\
    &\hspace{4mm}
    + 
    (1 - p_{0}) \int_{-\infty}^{\infty} \abs{\prox_{\frac{\alpha \lambda}{\beta \sqrt{\Delta}} f} (z)} p_{Z}(z) dz \\
    &= 
    p_{0} \cdot \frac{2\alpha}{\sqrt{\Delta}} \int_{\frac{\lambda}{\beta}}^{\infty} \paren{g - \frac{\lambda}{\beta}} p_{\txG}(g) dg \notag \\
    &\hspace{4mm}
    +
    (1 - p_{0}) \cdot 2 \int_{\frac{\alpha \lambda}{\beta \sqrt{\Delta}}}^{\infty} \paren{z - \frac{\alpha \lambda}{\beta \sqrt{\Delta}}} p_{Z}(z) dz, 
\end{align}
where $p_{Z}(z)$ is the PDF of the Gaussian distribution with zero mean and variance $1 + \frac{\alpha^{2}}{\Delta}$. 
The expectation in the second term of~\eqref{eq:Ex} can also be rewritten as 
\begin{align}
    &\Ex{\paren{\prox_{\frac{\alpha \lambda}{\beta \sqrt{\Delta}} f} \paren{X + \frac{\alpha}{\sqrt{\Delta}} G} - \paren{X + \frac{\alpha}{\sqrt{\Delta}} G}}^{2}} \notag \\
    &= 
    p_{0} \int_{-\infty}^{\infty} \paren{\prox_{\frac{\alpha \lambda}{\beta \sqrt{\Delta}} f} \paren{\frac{\alpha}{\sqrt{\Delta}} g} - \frac{\alpha}{\sqrt{\Delta}} g}^{2} p_{\txG}(g) dg \notag \\
    &\hspace{4mm} 
    + 
    (1 - p_{0}) \int_{-\infty}^{\infty} \paren{\prox_{\frac{\alpha \lambda}{\beta \sqrt{\Delta}} f} (z) - z}^{2} p_{Z}(z) dz \\
    &= 
    p_{0} 
    \paren{
        \frac{2 \alpha^{2} \lambda^{2}}{\beta^{2} \Delta} \int_{\frac{\lambda}{\beta}}^{\infty} p_{\txG}(g) dg 
        + 
        \frac{2 \alpha^{2}}{\Delta} \int_{0}^{\frac{\lambda}{\beta}} g^{2} p_{\txG}(g) dg
    } \notag \\
    &\hspace{2mm}
    + 
    (1 - p_{0}) 
    \paren{
        \frac{2 \alpha^{2} \lambda^{2}}{\beta^{2} \Delta} \int_{\frac{\alpha \lambda}{\beta \sqrt{\Delta}}}^{\infty} p_{Z}(z) dz 
        + 
        2 \int_{0}^{\frac{\alpha \lambda}{\beta \sqrt{\Delta}}} z^{2} p_{Z}(z) dz
    }. 
\end{align}
We can compute the above integrals by using 
\begin{align}
    \int_{a}^{b} p_{\txR}(r) dr 
    &= 
    P_{\txG}\paren{\frac{b}{\sigma_{\txr}}} 
    - P_{\txG}\paren{\frac{a}{\sigma_{\txr}}}, \label{eq:int_p} \\
    \int_{a}^{b} r p_{\txR}(r) dr 
    &= 
    \sigma_{\txr} 
    \paren{
        - p_{\txG}\paren{\frac{b}{\sigma_{\txr}}} 
        + p_{\txG}\paren{\frac{a}{\sigma_{\txr}}}
    }, \\
    \int_{a}^{b} r^{2} p_{\txR}(r) dr 
    &= 
    \sigma_{\txr}^{2} 
    \Biggl( 
        -\frac{b}{\sigma_{\txr}} p_{\txG}\paren{\frac{b}{\sigma_{\txr}}} 
        + \frac{a}{\sigma_{\txr}} p_{\txG}\paren{\frac{a}{\sigma_{\txr}}} \notag \\
    &\hspace{20mm}
        + P_{\txG}\paren{\frac{b}{\sigma_{\txr}}} 
        - P_{\txG}\paren{\frac{a}{\sigma_{\txr}}}
    \Biggr), \label{eq:int_w2p} 
\end{align}
where $p_{\txR}(r)$ is the PDF of the Gaussian distribution with zero mean and variance $\sigma_{\txr}^{2}$. 
\ifCLASSOPTIONcaptionsoff
  \newpage
\fi
%
%------------------
% References
%------------------
\bibliographystyle{IEEEtran}
\bibliography{MyBib, others}
\begin{IEEEbiographynophoto}{Ryo Hayakawa}
received the bachelor's degree in engineering, the master's degree in informatics, and Ph.D.\ degree in informatics from Kyoto University, Kyoto, Japan, in 2015, 2017, and 2020, respectively. 
He is currently an Assistant Professor at Graduate School of Engineering Science, Osaka University. 
He was a Research Fellow (DC1) of the Japan Society for the Promotion of Science (JSPS) from 2017 to 2020. 
From 2023, he is an Associate Editor of IEICE Transactions on Fundamentals of Electronics, Communications and Computer Sciences. 
He received the 33rd Telecom System Technology Student Award, APSIPA ASC 2019 Best Special Session Paper Nomination Award, and the 16th IEEE Kansai Section Student Paper Award. 
His research interests include signal processing and mathematical optimization. 
He is a member of IEEE and IEICE. 
\end{IEEEbiographynophoto}
\vfill
\end{document}